\documentclass[printer]{aa}
\usepackage{color}
\usepackage{soul}
\setulcolor{red}
\usepackage{txfonts}
\usepackage{natbib}
\usepackage{graphicx}

\begin{document}

\title{Counter-dispersed slitless-spectroscopy technique:
planetary nebula velocities in the halo of NGC 1399 \thanks{Based on observations made with the VLT at Paranal Observatory under programme ID 066.B-0278}}
\author{E. K. McNeil\inst{1,2,3} \and M. Arnaboldi\inst{2} \and K.C. Freeman\inst{1} \and O.E. Gerhard\inst{3} \and  L. Coccato\inst{3} \and P. Das\inst{3}}

\institute{Mt Stromlo Observatory, Australian National University, Cotter Road, Weston Creek, ACT 2611, Australia;
E-mail: emcneil@mso.anu.edu.au \and European Southern Observatory, Karl-Schwarzschild-Str. 2, D-85748 Garching, Germany; \and Max Planck Institute for Extraterrestrial Physics,
Karl-Schwarzschild-Str. 1, 85741 Garching, Germany;}

\date{Received date: 13 March 2010  / 
Accepted date: 04 May 2010}

\abstract    
   {}
   {Using a counter-dispersed slitless spectroscopy technique,
  we detect and measure the line-of-sight velocities of 187 planetary
  nebulae (PNe) around one of the nearest cD galaxies, NGC 1399, with
  FORS1 on the VLT. }
   {We describe the method for identifying and
  classifying the emission-line sources and the procedure for
  computing their J2000 coordinates and velocities. The number of PN
  detections and the errors in the velocity measurements (37 km
  s$^{-1}$) indicate that this technique is comparable to other
  methods, such as that described by Teodorescu et al.\ (2005).}
   {  We present the spatial distribution of the PNe and a basic
  analysis of their velocities.  The PN two-dimensional velocity field
  shows marginal rotation consistent with other studies. We also
  find a low-velocity substructure in the halo and a flatter
  velocity-dispersion profile compared to previous observations that extends to $\sim$400\arcsec. The
  detection of a low-velocity subcomponent underscores the importance
  of discrete velocity tracers for the detection of un-mixed
  components. }
	{The new velocity-dispersion profile is in good agreement
  with revised velocity dispersions for the red globular clusters in
  NGC 1399, using the data of Schuberth et al.\ (2009).  The outer
  parts of this profile are consistent with one of the dynamical
  models of Kronawitter et al.\ (2000), which corresponds to a
  circular velocity of $\simeq$340 km s$^{-1}$ and a rescaled B-band
  mass-to-light ratio of $\simeq$20 at 7' radius.  These measurements 
  trace the kinematics of the outer halo and disentangle the heterogenous 
  populations in the Fornax Cluster core. The new data set
  the stage for a revised dynamical model of the outer halo of NGC
  1399.}


\titlerunning{Counter-dispersed spectroscopy of NGC 1399}
\authorrunning{McNeil \it{et al.}}

\maketitle
\section{Introduction}

An important key to understanding the formation and evolution of galaxies is to examine the kinematics of the halo where the orbits have long dynamical timescales and the mass profile is dominated by dark matter. In disk galaxies, HI measurements generally provide line-of-sight velocities out to the edges \citep{Rubin:1977,Bosma:1978}. Measuring kinematics in the outer parts of early-type galaxies is more complicated since HI measurements are usually unavailable, and the signal-to-noise of the integrated light spectra drops off before the mass profile flattens. Specifically, long-slit spectra provide measurements for  the first 2-3 \textit{R$_{e}$} \citep{Kronawitter:2000}, multi-object spectroscopy provides line-of-sight velocity distributions out to 3 $R_{e}$ \citep{Proctor:2009} and  IFUs can take measurements at 4 $R_{e}$ \citep{Murphy:2007,Weijmans:2009}. Discrete tracers like planetary nebula  provide the necessary kinematic information to examine the regions beyond these limits and learn about the formation history of the outer regions of elliptical galaxies \citep{Hui:1995, Arnaboldi:1996}. 

The planetary nebula (PN) stage occurs at the end of a star's life when the atmosphere has been ejected and is subsequently ionized by UV radiation from the exposed stellar core. Due to the extreme rarefaction of the gas in the atmosphere, these objects emit up to 15\% of their light at [OIII] ($\lambda$5007) \citep{Dopita:1992}, which allows their detection and velocity measurement out to a distance of 100 Mpc \citep{Gerhard:2005}. Moreover, the PN number density traces the light in early-type galaxies \citep{Coccato:2009}. Because they follow the same spatial distribution as stars,  PNe provide kinematics that are directly related to integrated light measurements. We are able to use planetary nebulae as kinematic tracers out to more than 7 $R_e$ for measuring the dark matter distribution and angular momentum content in the halo  \citep{Mendez:2001, Romanowsky:2003, Douglas:2007, De-Lorenzi:2008, De-Lorenzi:2009, Napolitano:2009, Coccato:2009}.

 In general, globular clusters follow a different spatial distribution so their kinematics do not trace the same population as the stars \citep{Harris:1991, Ashman:1998, Brodie:2006}.  The globular cluster system is dominated by the blue component of a bimodal color distribution in most galaxies. The radial distribution of this component is flatter than the integrated light from the stars. However, in many galaxies including NGC 1399, the red globular cluster sample does follow the integrated light \citep{Brodie:2006, Schuberth:2009}, and we expect the kinematics of this subset to be comparable to those of the integrated light and the PNe.

Traditionally, PNe are detected using an on-band/off-band technique \citep{Ciardullo:1989}. The sky is imaged through a narrow-band filter centered at the redshifted [OIII] $\lambda$ 5007 line and then again in a broad band that does not contain any bright emission lines. PNe are too faint to detect in the broad-band image, but they appear in the narrow-band image. Foreground stars appear in both narrow- and broad-band images, so the wide filter is used to distinguish them from the PNe.  Once the positions are measured from the on-band image, the velocities are then typically obtained using multi-slit or fibre spectroscopy \citep{Arnaboldi:2004, Doherty:2009}. 

More recently, a number of techniques have been developed that allow detection and velocity measurement to be done in the same observing run. \citet{Mendez:2001} identify PNe using narrow- and broad-band images, but then follow with a dispersed image in the same run. The velocity of each PN can be calculated from the difference between the object's position in the narrow-band frame and the dispersed frame. This technique has also been applied to NGC 1344, another Fornax galaxy, by \citet{Teodorescu:2005}. We compare the efficiency of those observations to ours in \S \ref{mendez}.

\begin{figure}
\includegraphics[width=0.45\textwidth]{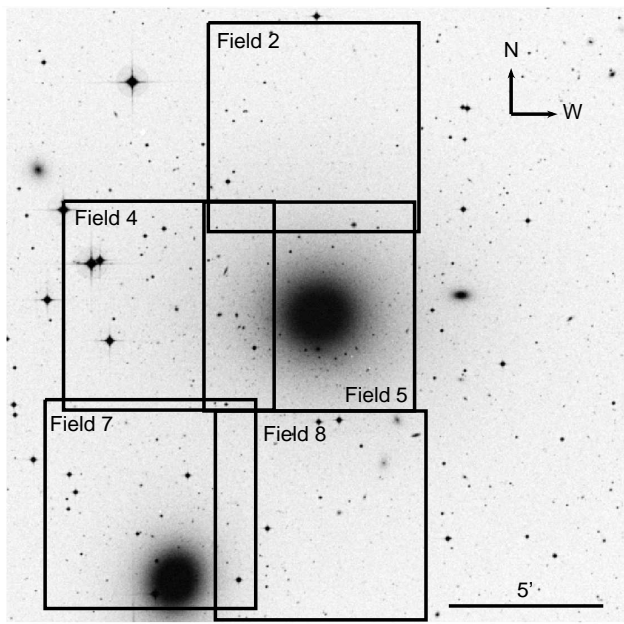}
\caption{ \label{grid} DSS image of NGC 1399 overlaid with the five observed fields. The fields were intentionally overlapped in order to confirm consistent astrometry and velocity calibration across the observations. NGC 1404 is seen in field 7.  }
\end{figure}

The PN Spectrograph (PN.S) is an instrument on the William Herschel Telescope dedicated to measuring PN velocities in nearby galaxies \citep{Douglas:2002}. The pupil is split in half before being dispersed in opposite directions in twin spectrographs. A combination of the two exposures, which are simultaneously generated, allows identification of emission-line objects and the measurement of their velocities using the separation between the positions in the two spectral images. 

For extremely faint samples in distant galaxies, observing the entire field through slits significantly reduces the signal from the sky. Multislit Imaging  Spectroscopy involves a grid of slits that are stepped until the entire field has been spectrally imaged.  These types of observations have detected samples of intracluster PNe 100 Mpc away \citep{Gerhard:2005, Arnaboldi:2007, Ventimiglia:2008}. 

The counter-dispersed slitless technique in this paper uses only two exposures for each field to obtain positions and velocities of PNe in the cD galaxy, NGC 1399.  The field is first observed with a dispersed image. Next the spectrograph is rotated 180 degrees and the same field is exposed again, this time with the dispersion in the opposition direction. As in the PN.S observations, the velocity is a function of the separation between the position of the PN in the two frames.   In this way, the slitless technique avoids the two-stage selection and measurement process. Because there are no slits or fibres, the light loss is reduced. The number of measurable PN velocities is not limited by the number of available fibres or the restrictive geometry of the slits. 

 NGC 1399 is a cD galaxy in the Fornax cluster \citep{Schombert:1986}. It is one of the closest cD galaxies and therefore promises one of the best kinematic data sets for the outer halo. Located near the center of their clusters, cD galaxies are believed to form through accretion and mergers with other cluster members. As such, these galaxies are valuable clues to the formation history of galaxies and the evolution of clusters \citep{Hausman:1978}.

This paper describes the latest kinematic data for the outer parts of the galaxy. It is part of a larger project intended to dynamically model the galaxy using PNe and X-ray constraints. NGC 1399 is well studied with kinematic measurements from long-slit spectra \citep{Saglia:2000}. Previous to this work, the radial extent of the PN kinematics was limited to the velocities from \citet{Arnaboldi:1994a} at 24kpc. The current sample measures more than 4 times as many PN velocities out to more than twice the radius. The resulting kinematics provide a significant improvement in understanding the velocity field of the halo. 

\begin{figure*}
\includegraphics[angle=270, width=0.9\textwidth]{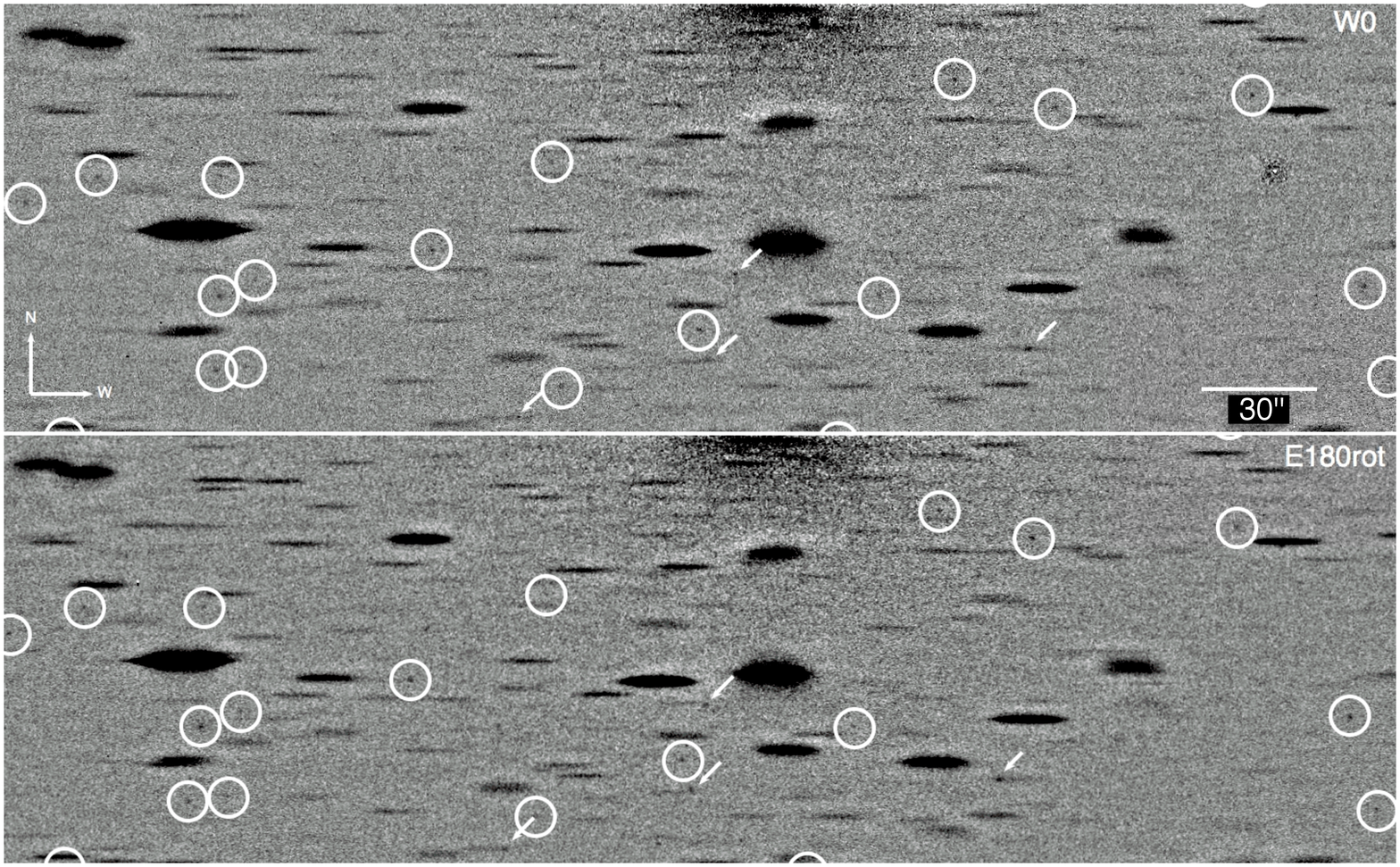}
\caption{ \label{cdipic} Part of the central field (field 5). The top frame is the W0 exposure and the bottom frame is the counter-dispersed E180 exposure after rotating back in the computer. The diffuse light at the top center of each frame is the southern tip of the bright central region of NGC 1399. In these images, the stars appear as streaks in the $x$-direction (the direction of dispersion). Examples of planetary nebula candidates are circled. Notice that there is a small difference in the PNs' $x$-positions in the W0 and E180 fields due to their velocity. The arrows point to emission-line objects that are associated with some continuum-- these are background galaxies.}
\end{figure*}

\begin{figure*}
\includegraphics[width=0.9\textwidth]{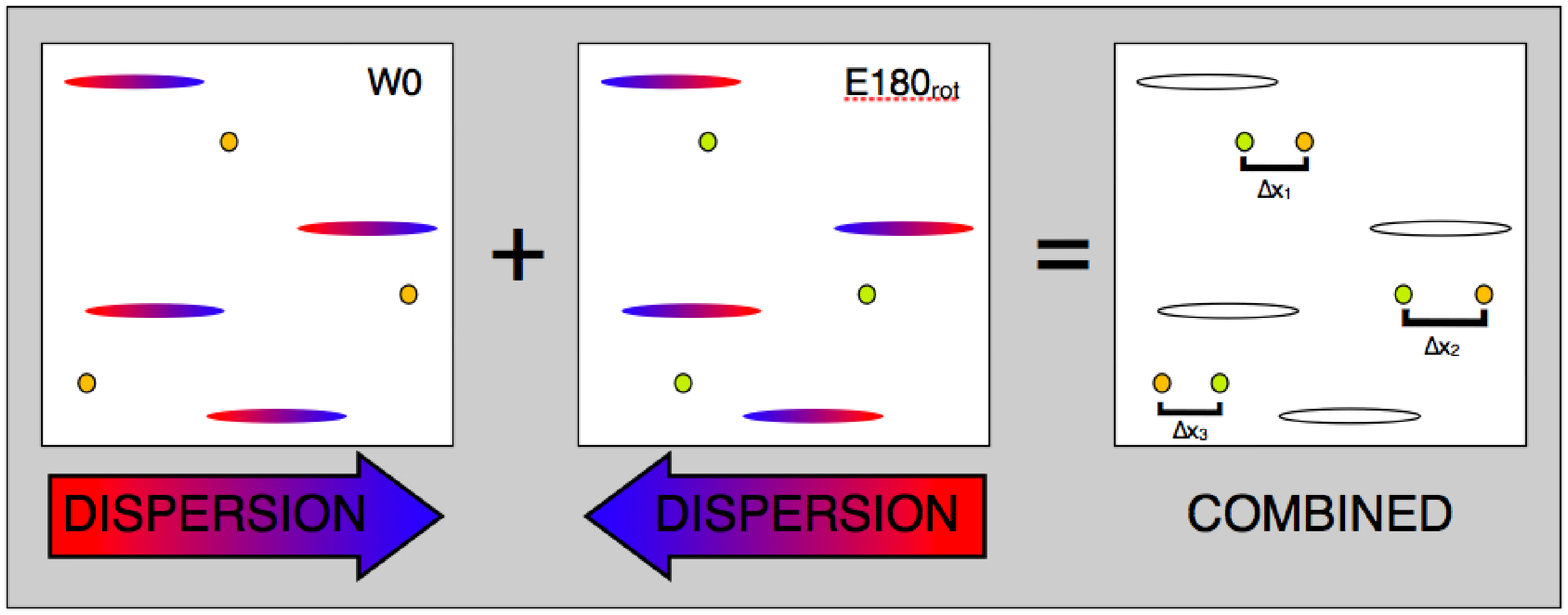}
\caption{ \label{cdicartoon} Schematic describing the counter-dispersed imaging technique. In the field shown, there are four stars (shown as rainbow continuum sources) and 3 planetary nebulae (shown as small emission-line dots). The first frame shows the W0 field with dispersion to the right. The second frame, E180$_{rot}$, shows the 4 stars in the same locations but now dispersed in the opposite direction. This image was taken at P.A.= 180$^{\circ}$ and rotated back during the reduction. Notice that the PN change position between these frames because of their velocity. We combine the dispersed and counter-dispersed images in the final frame by aligning the stars. The separation ($\Delta x$) of each pair of PN detections depends on the PN's velocity.  }
\end{figure*}
We describe our observations in \S 2 before moving on to the data reduction, the image registration and the detection of planetary nebula in \S3. We discuss the measurement of the PN velocities in \S 4, and \S 5 addresses the technique for mapping the astrometry on spectral images. The analysis of the sample, including discussions of objects associated with NGC 1404 and a low-velocity subcomponent, is covered in \S6. We explore the kinematics of the halo of NGC 1399 in \S 7 and give a summary of this work in \S8. Throughout the paper, we assume the distance from \citet{Tonry:2001} of 19.95 Mpc,  which means that 1\arcsec \ corresponds to 96.7 pc.
\section{Observations}
\subsection{Instrumental set-up}
Observations of five fields around NGC 1399 were carried out with FORS1 \citep{Appenzeller:1998} on the 8-m VLT at Paranal Observatory during the nights of 25-27 Nov 2000. The five fields represent a subset of the 3$\times$3 grid proposed, which was centered on the galaxy as shown in Fig.~\ref{grid}. Information from the observing log is tabulated in Table~\ref{olog}.  The FORS1 field-of-view is $6'.8\times6'.8$ or 39.5kpc on each side at 19.95 Mpc \citep{Tonry:2001}.   The spatial scale is $0''.25/$pixel.

We used the 600B grating with a measured mean dispersion of 1.19\AA \ pix$^{-1}$. The interference filter, FILT\_503\_5+86, is centered near the wavelength of the redshifted [OIII] emission at the systemic velocity of NGC 1399 (1425 km s$^{-1}$, \citealt{Graham:1998}). The filter, with a FWHM of 60\AA, limits the length of the spectra of continuum sources which would otherwise cover much of the field. It allows detection of PNe with velocities within $\pm$1800 km s$^{-1}$ of the systemic velocity. For some fields, bright stars were masked out.

With this instrumental configuration, the direction of dispersion is along the horizontal axis ($x$-axis) depicted in Fig.~\ref{cdipic}. Because of the slitless technique, both axes have spatial information. In the focal plane, the presence of the grism causes an anamorphic distortion resulting in a contraction in the direction of dispersion (see discussion in \S \ref{calibrations}).

\begin{table}
\caption{Observing log$^*$. }
\label{olog}
\centering
\begin{tabular}{c l l c c}
\hline \hline
 Field & P.A. Label & Exp. Time & Seeing\\
 && (s) & (FWHM)\\
\hline
NGC1399-2 & W0 & 3 $\times$ 1200 & $1.42^{\prime\prime}$\\
NGC1399-2 & E180 & 3 $\times$ 1200 & $1.34^{\prime\prime}$\\
NGC1399-4 & W0 & 3 $\times$ 1200 & $1.23^{\prime\prime}$\\
NGC1399-4 & E180 & 3 $\times$ 1200 &  $1.10^{\prime\prime}$\\
NGC1399-5 & W0 & 3 $\times$ 1200 & $1.11^{\prime\prime}$\\
NGC1399-5 & E180 & 3 $\times$ 1200 & $1.09^{\prime\prime}$\\
NGC1399-7 & W0 & 3 $\times$ 1200 &  $1.04^{\prime\prime}$\\
NGC1399-7 & E180 & 3 $\times$ 1200 &  $1.22^{\prime\prime}$\\
NGC1399-8 & W0 & 3 $\times$ 1200 & $1.09^{\prime\prime}$\\
NGC1399-8 & E180 & 3 $\times$ 1200 &  $1.10^{\prime\prime}$\\
\hline
\end{tabular}
\begin{list}{}{}
\item[$^{\mathrm{*}}$]  Each field is observed at P.A.=0$^{\circ}$ and 180$^{\circ}$, \ labeled W0 and E180 respectively.
\end{list}
\end{table}

\subsection{Observing mode}
\label{cdi}
In this technique, the positions and velocities are obtained using two narrow-band, dispersed exposures of the same field. In the first, an image is acquired at position angle (P.A.) 0$^{\circ}$ (W0). The result is that stars become streaks in the direction of dispersion with a length determined by the dispersion and the width of the filter. Emission-line sources, such as PNe, appear as dots whose width is related to the seeing in $y$-direction and the linewidth and the seeing in the $x$- (wavelength) direction.  The location of the PN on the spectral image is a function of its position on the sky, its velocity, and the instrumental distortion. See Fig.~\ref{cdipic} for an example image. In the second exposure (E180), the instrumental set-up remains the same except that the spectrograph is rotated  to P.A. = 180$^{\circ}$. In this configuration, the sky appears rotated 180$^{\circ}$ on the plane of the CCD--- in order to register the two images we must counter-rotate the E180 frame during the software reduction. See Fig.~\ref{cdicartoon} for an illustration of the technique. The velocity is related to the difference in position between the PN in W0 and the counter-rotated E180 frame (E180$_{rot}$), after the images have been registered using the stars as shown in the final frame of Fig.~\ref{cdicartoon}.

\subsection{Calibrations}
\label{calibrations}
The position of each PN in the dispersed image is a combination of its position on the sky, its line-of-sight velocity and the distortion from the instrument; we must quantify and separate these contributions in order to measure the velocity. The grism introduces an anamorphic distortion into the field that needs to be corrected before we measure velocities.  Also, the dispersion changes as a function of position and the filter bandpass is dependent on the location in the field.  

We used three sets of daytime calibration frames based on a $10\times19$ grid of slitlets imaged through the narrowband [OIII] filter. In the first set, the slitlets are imaged in undispersed white light. This gives us the positions of the slitlets in the grid. In the second set, the white light is dispersed. This set determines the filter bandpass as a function of position on the CCD. The final set is illuminated with a dispersed HeAr lamp. There are three lines covered by our bandpass: $\lambda 5015.67,   \lambda 5047.74$ and $\lambda 5085.82$. This frame is used to evaluate the anamorphic distortion and to measure the local dispersion. 

\begin{figure*}
\includegraphics[width=0.95\textwidth]{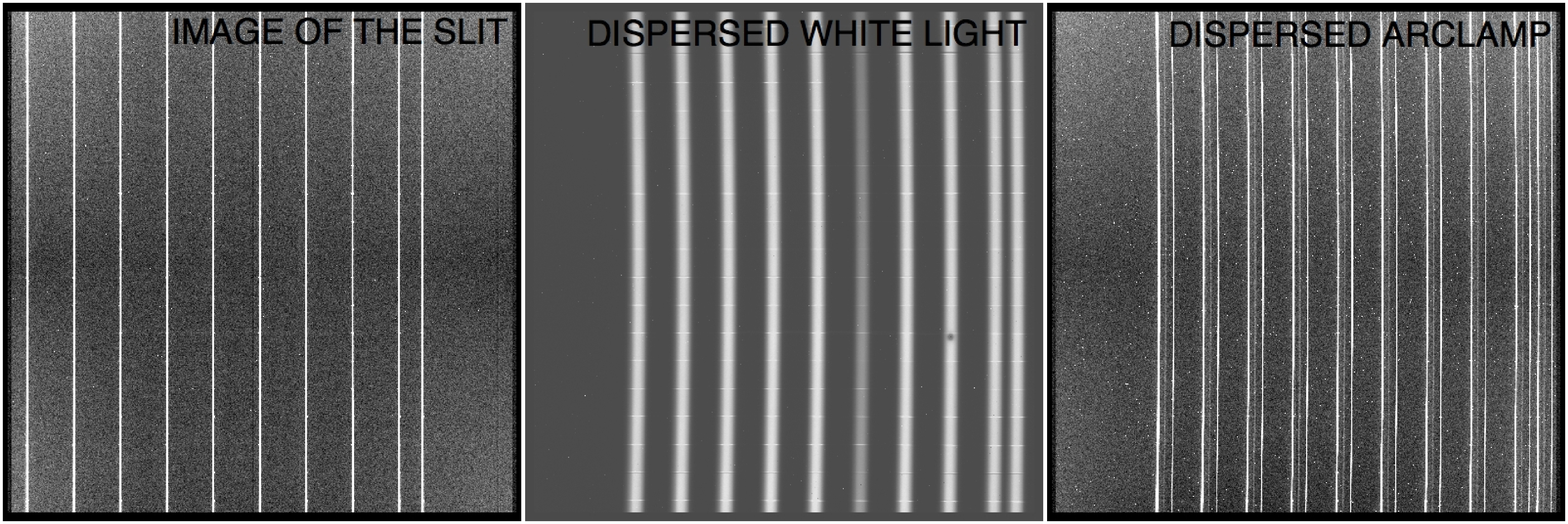}
\caption{ \label{calframes} Calibration frames. Each vertical line is comprised of 19 nearly-adjacent slitlets. Using this $10 \times 19$ grid of slitlets, we imaged the grid (left), added a grism to the light path to disperse the white light (center) and illuminated the 190 slitlets with a HeAr lamp and dispersed the light (right). In combination, these three frames allow us to map the anamorphic distortion, measure the bandpass shift as a function of CCD position and calculate the local dispersion.  }
\end{figure*}

The daytime calibration exposures for the following steps were taken over three days where each slitlet position was exposed once in each configuration (image of the slitlet, dispersed white-light spectrum, and dispersed HeAr spectrum). The slitlets are formed by 19 pairs of moveable fingers, each calibrated at 10 positions across the field. All of the positions were co-added to create a master white-light image, a master white-light spectrum, and a master  HeAr lamp spectrum. The three master calibration frames are shown in Fig.~\ref{calframes}. The image processing and analysis necessary for these calibrations was done in IRAF and Mathematica. The $y$-position is taken to be the vertical center of each slitlet.  To measure the position in the dispersion direction,  40 rows were summed along the slitlet and a Gaussian was fit--- the $x$-position is taken to be the mean of the Gaussian. See \S \ref{veloffset} for a discussion of how well it is possible to measure the central wavelength of the filter bandpass. 
\begin{figure}
\includegraphics[width=0.45\textwidth]{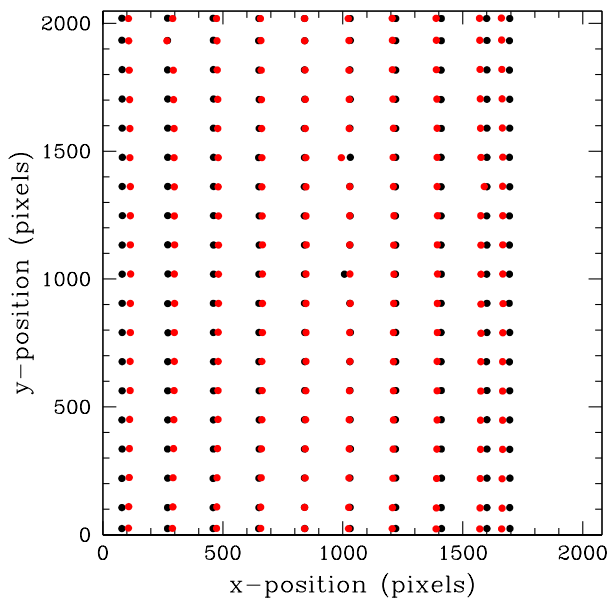}
\caption{ \label{anamorphic} Anamorphic distortion. Each black point represents the location of the midpoint of a slitlet in the undispersed image frame. The red points are the corresponding slitlets as measured using the $\lambda$5015.67 line in the dispersed HeAr lamp. The entire red grid has been shifted by the mean offset in order to display higher order distortions rather than the rigid shift caused by the disperser. Notice that the grism has introduced a sort of pincushion effect with the largest effects being at the right and left edges towards the middle. }
\end{figure}

\subsubsection{Mapping the anamorphic distortion} \label{anam}The first calibration step is to use the $x$,$y$-positions from the image of the slitlets in the first calibration frame and the positions of the brightest line ($\lambda 5015.67$) in the third calibration frame to map the anamorphic distortion introduced by the grism. With no distortion, the HeAr line would show a constant offset from the image of the slitlet on the CCD. We tabulate the difference in $x$ and $y$ and use it to make a map of the anamorphic distortion using the IRAF task \texttt{geomap}. A graphical representation of the matrix is shown in Fig.~\ref{anamorphic}. All 190 slitlets shown were used  in a 5th-order polynomial fit. Using  \texttt{geotran}, we applied the solution from \texttt{geomap} to all of our images to remove the distortion. 

Effectively, this transformation converts the data from the spectral into the image plane. We define the spectral and image planes like \citet{Arnaboldi:2007};  the spectral plane is the dispersed image as observed, and the image plane has been corrected for the distortion introduced by the grism thereby making an undispersed, distortion-corrected image for one wavelength ($\lambda 5015.67$). The PN identification and velocity measurement were done on the image-plane images because they are distortion-corrected. 

\subsubsection{Calculating the local dispersion}
Next, we measure the distance between the 5015.67\AA\ and 5085.82\AA\ lines in the dispersed HeAr lamp image to derive the local dispersion. We assume that the dispersion is linear on this scale. The two-dimensional,  2nd-order function describing the dispersion at each of the 190 calibration slitlets is fit with \texttt{surfit}, and we interpolate to get a measurement of the local dispersion at any location.

Because we measure the dispersion in the non-distortion-corrected (spectral) calibration frame, we multiply the local dispersions by the linear term from the \texttt{geomap} solution in \S \ref{anam} when we apply them to relate position to wavelength.  This is possible because the higher-order terms are small compared to the linear term of the anamorphic-distortion fit. This converts our dispersions from the measured spectral-plane dispersions to image-plane dispersions. 

\subsubsection{Measuring the bandpass shift} \label{secbpshift}
\begin{figure}
\includegraphics[width=0.35\textwidth, angle=270]{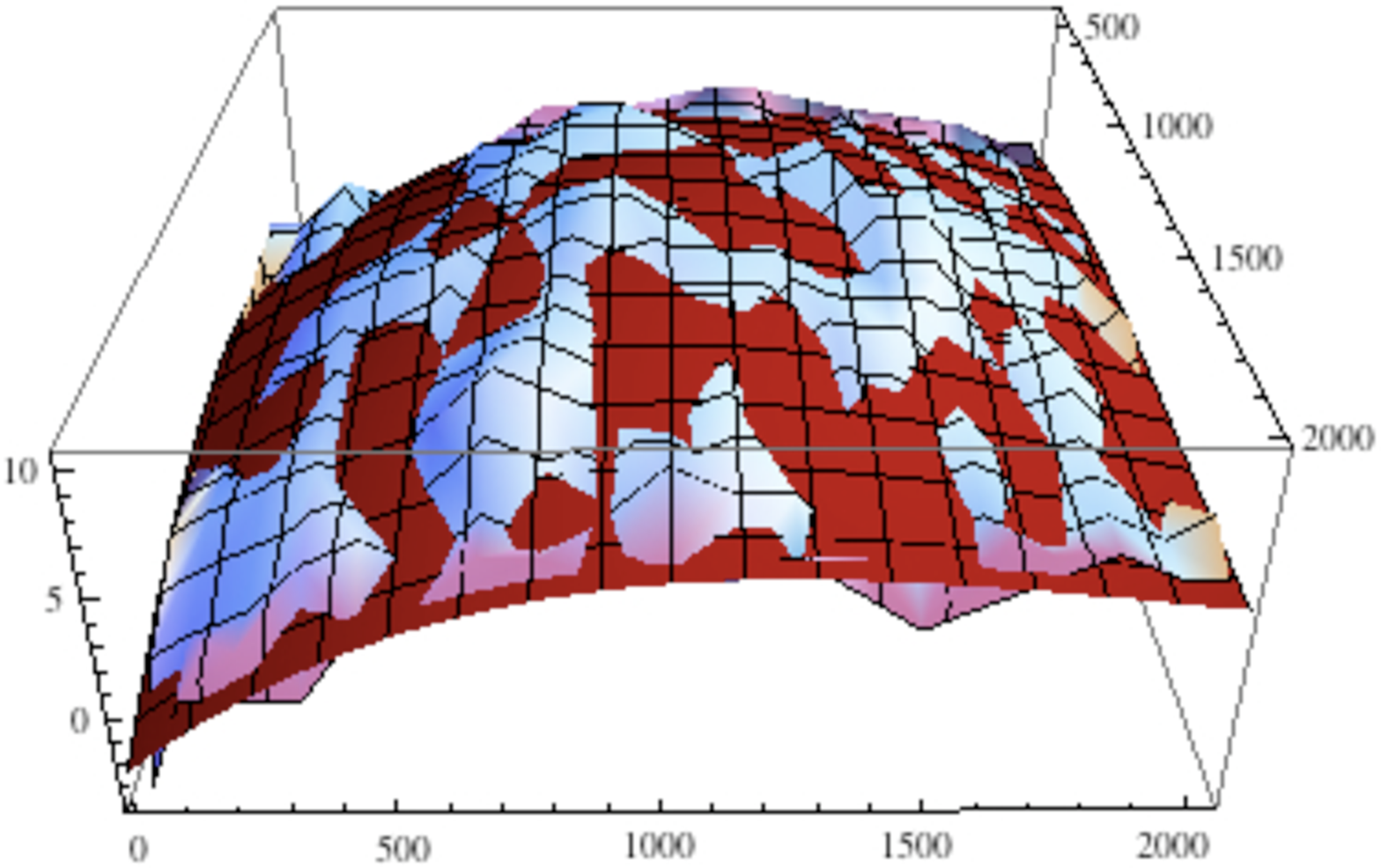}
\caption{ \label{bpshift} Bandpass shift as a function of position on the CCD.  The shift is measured by comparing the measured center of the bandpass to the location of the $\lambda$5015.67-line at 190 locations across the CCD. The bumpy, blue-toned surface represents the measured shift. The smooth red surface is a 2-D cubic fit to the blue data. We evaluate the fit when correcting for this shift in the stars. All three axes are measured in pixels.}
\end{figure}

Finally, we map the bandpass shift as a function of position. We do this by comparing the Gaussian center of the dispersed white light to the position of the $\lambda$5015.67 line.  We expect there to be an underlying smooth, 2-D function that describes the bandpass shift caused by the presence of an interference filter at a small inclination in a converging beam.  We modeled this function with a 2-D cubic using Mathematica and now we can evaluate the bandpass at any location on the CCD.  Fig.~\ref{bpshift} shows the measured bandpass shift across the CCD in light blue. The smooth red surface represents the model for the underlying smooth function that determines the local bandpass. The rms of the residuals is less than 1 pixel. 

When we align the W0 and E180 images using the stars, we must first correct for the bandpass shift. In the W0 and E180 images of the same star, the light path is different causing different bandpass shifts in the two images. In general, the location of a star in our images is measured with a 2-D Gaussian using the \texttt{n2gaussfit} task in IRAF. For each star used in the alignment, we evaluate the two-dimensional cubic function at that location and apply the shift to the \texttt{n2gaussfit} coordinates. The bandpass shift is measured in the spectral plane, but we want to register image-plane images. Therefore, using \texttt{geoxytran}, we apply the anamorphic distortion correction from \texttt{geomap} to obtain the distortion-corrected bandpass-shifted location of each star. These new coordinates are used to obtain the rigid shift that will be applied to the distortion-corrected images before calculating the velocity of the PNe. Because we measure the center of the filter relative to the $\lambda$5015.67  line, it becomes our fiducial wavelength. The E180 and W0 images will be aligned on the $x,y$ positions of this wavelength in the stars, and the wavelength of the PNe emissions will be measured relative to that line.

\section{Data reduction, registration and identification of emission-line sources}
\subsection{Data Reduction}
The data reduction for the counter-dispersed technique was carried out in IRAF. Each field is imaged with three-1200s exposures at each position angle.  The frames are bias subtracted and flat fielded individually. The sky flats  were taken at twilight using the narrow-band filter and the grism. The three frames at each P.A. were combined using \texttt{imcombine} with a standard sigma-clipping for cosmic-ray removal. 

The result is a total of 10 combined, bias-subtracted, flat-fielded frames-- 5 at P.A.=W0 and 5 at P.A.=E180. We subtract the background from each frame by smoothing with \texttt{fmedian} and subtracting the smoothed background from the spectral image.  Next each of the frames is corrected for anamorphic distortion using IRAF task \texttt{geotran} as described in \S \ref{anam}-- this transforms the frame from the spectral to the image plane. We rotate the E180 image by 180$^{\circ}$ so that the stars are in comparable positions in the W0 and E180$_{rot}$ frames; the calibration frames are no longer applicable to E180$_{rot}$ because the direction of dispersion is opposite. 

\subsection{Registering the images}
We must now register the W0 and E180$_{rot}$ images in order to measure the positions and velocities of the PNe. The registration of the W0 and E180$_{rot}$ frames is done with a rigid shift. Note that the shift is calculated from stars, which need to be adjusted for the bandpass shift as described in \S \ref{secbpshift}. More specifically, the frames are aligned on the virtual position of a set of $\sim 10$ monochromatic stars at $\lambda 5015.67$. To achieve this, we measure the locations of the stars in the spectral plane, apply the bandpass shift, correct for anamorphic distortion for the shifted coordinates using \texttt{geoxytran}, rotate the distortion-corrected, shifted positions from E180 only, and calculate the shift that will be applied to the images. After the registration of the frames, we can create a difference image and identify emission-line objects. The resulting pair of registered images looks like Fig.~\ref{cdipic}.

\subsection{Identification of emission-line objects}
In these images, emission-line sources are easily distinguished from stars because they appear as points instead of streaks. For the identification we rely on the W0, E180$_{rot}$ and W0--E180$_{rot}$ difference images of each field. Specifically, an object is added to the PN catalog if it is point-like in both the spatial and dispersion directions, and if it appears in both W0 and E180$_{rot}$ at the same $y$-position and with similar intensity (it need not be exact since the seeing may vary between the two images). 

\begin{figure}
\includegraphics[width=0.45\textwidth]{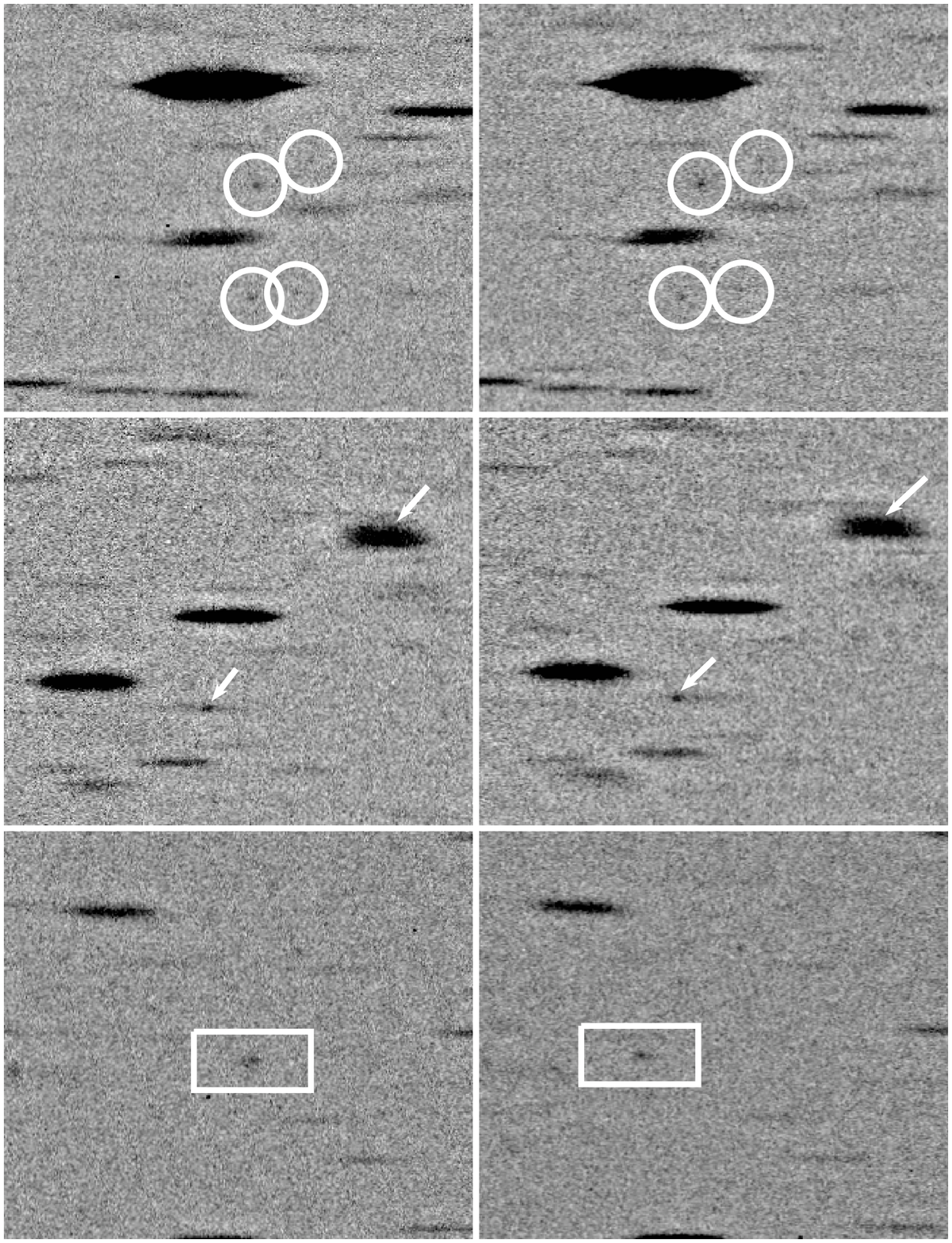}
\caption{ \label{stamp} Examples of the three classes of emission-line objects identified in our sample. Each frame is 1\arcmin.5 across and north is up. {\bf top:} Four PN are circled. They are point-like in both directions. {\bf middle:} The arrows point to two examples of background galaxies. The top-right one is a continuum source that is extended in the spatial (vertical) direction. The bottom-left one shows an emission line superimposed on some continuum. This could be a Lyman-$\alpha$ galaxy. {\bf bottom:} The object in the box is slightly extended in the wavelength direction-- we are seeing the double emission of an [OII] emitter. }
\end{figure}

We identify three kinds of emission-line objects: 

\begin{itemize}
\item Those that are point-like in both wavelength and space-- these are planetary nebula candidates;
\item Those emission-line sources that are associated with a continuum-- these are background galaxy candidates including some Lyman-$\alpha$ galaxies at z=3.1. These objects can be either point-like or extended in the spatial direction;
\item Those with resolved or double emission in wavelength. They can be extended or point-like in the spatial direction. These are also background galaxies and include some [OII] emitters. 
\end{itemize}

Examples of these classes of emission-line sources are shown in Fig.~\ref{stamp}. Further examples are available in Fig.~\ref{cdipic}. We identify a total of 411 emission-line objects in our fields.  The breakdown of these detections into types of emission-line objects is shown in Table~\ref{breakdown}. The definitions used to distinguish members of NGC 1399 from those objects belonging to NGC 1404 and the low-velocity subcomponent are described in \S \ref{separation} and \S \ref{lowvelpop} respectively. For the planetary nebulae, we can use the known rest wavelength of the emission, $\lambda$5006.8, to acquire the velocities using measured separation between the W0 and E180 frames-- this is described in \S \ref{velcaleqn}.

\begin{table}
\caption{ Breakdown of emission-line objects}
\label{breakdown}
\centering
\begin{tabular}{ c c c c c}
\hline \hline
  NGC 1399 & NGC 1404  & Low-vel.  & Unassigned & Background\\
  PNe & PNe & PNe & PNe& Galaxies\\
\hline
 146 & 23 & 12 & 6 & 224 \\
\hline
\end{tabular}

\end{table}

\subsubsection{The efficiency of the technique}
\label{mendez}
We found a total of 187 PNe in the five fields, including 146 bound to NGC 1399. As a point of comparison, \citet{Teodorescu:2005} applied a related technique to NGC 1344, another Fornax galaxy, using the VLT  and found 197 PNe. In their technique (summarized in \citet{Mendez:2001}), each field is imaged with broad- and narrow-band filters, and then a grism is added and the narrow-band image is dispersed. The velocity would be measured from the separation between the PNe in the narrow-band image and the dispersed, narrow-band image. 

As elliptical galaxies in the same cluster, we expect the PN detection efficiencies in these two galaxies to scale with their color and luminosity.  NGC 1399 is red compared to NGC 1344 (B-V=0.95 versus B-V=0.87  based on the extinction-corrected total luminosities published by \citet{de-Vaucouleurs:1991}). According to \citet{Hui:1993}, the difference in color would indicate that the luminosity specific PN number density of NGC 1399 is only 43\% of that of NGC 1344.  Also NGC 1399 is almost a full magnitude brighter which should have the effect of nearly doubling the number of PN detected in NGC 1399. Finally, Teodorescu et al. covered the area around NGC 1344 uniformly, while the NGC 1399 observations only covered 5/9ths of the proposed grid. We can estimate the number of detections we are missing by symmetrizing the detections from the fields we do have. This would increase our sample by about 55 PNe.

 If we adjust our detection rate for the difference in color, magnitude and coverage, we would have detected 224 PNe, which is similar to the 197 NGC 1344-bound objects found by \citet{Teodorescu:2005}.  We conclude that the counter-dispersed spectroscopy technique described in this paper provides an efficient strategy for collecting large PN velocity samples in the local volume and is comparable to the singly-dispersed imaging technique \citep{Mendez:2001, Teodorescu:2005}. 

\section{Planetary nebula velocities}

\subsection{Velocity calibration} 
\label{velcaleqn}
The data reduction and the registration were done using the positions of the stars. As continuum sources, their spectra are approximated by the nearly-symmetrical filter curve (see discussion in \S \ref{veloffset}) and adjusted to compensate for the filter bandpass shift. Unlike stellar continuum sources, planetary nebulae are not subject to the filter bandpass shift and their positions on the CCD are dependent on both their positions in the sky and their velocities. The counter-dispersion causes each PN to appear in different places in the W0 and E180$_{rot}$ relative to the stars (i.e. to the virtual position of monochromatic stars). This principle is illustrated in Fig.~\ref{cdicartoon}.  The velocity is a function of the distance between the position of the PN in the W0 and E180 exposures and offset by the wavelength we selected for the stellar alignment ($\lambda 5015.67$). More specfically, 

\begin{equation}
	\lambda=\lambda_0 + \frac{d\lambda}{dx} \frac{\Delta x}{2},
\label{deltaX2lambda}
\end{equation}

\noindent
where $\lambda$ is the measured wavelength of the planetary, $\lambda_0$ is a fiducial wavelength determined by the registration procedure (in this case, $\lambda 5015.67$), $d\lambda/d$x is the local dispersion and $\Delta x$ is the separation, in pixels,  between the PN locations in the W0 and E180 fields.

Using this calibration, we generate line-of-sight velocity measurements for the PNe in our sample. The mean measured velocity is different from the systemic velocity of the galaxy--- this indicates a zero-point offset in the calibration which is discussed in \S \ref{sysoff}. First, in \S \ref{verr} we will determine the errors. In \S \ref{caveat}, we seek to demonstrate that the velocities are internally consistent and therefore represent an accurate 2-D relative-velocity field. 

\subsection{Errors on the velocity measurements}
\label{verr}
The individual velocity errors are a combination of the errors in reduction, registration and measurement. The total error is calculated using 17 duplicates-- these are PN that are detected in the overlap region between two fields. For each set of two fields, the mean difference in the duplicates is estimated (see section \ref{lock}) and the individual velocity errors for our sample are determined from the standard deviation of the differences between all the pairs from overlapping areas.  We measure an rms in the difference of 52 km s$^{-1}$, corresponding to an individual velocity error of 37 km s$^{-1}$. This value is comparable to the 40 km s$^{-1}$ error on the velocities in \citet{Mendez:2001}, but it is larger than the 20 km s$^{-1}$ errors in the Planetary Nebula Spectrograph data \citep{Douglas:2002}.

\subsection{Testing the internal consistency of the velocity calibration}
\label{caveat}
Due to the complexity of the calibrations for counter-dispersed imaging, we seek to confirm that our velocity calibration is correct. The following three tests confirm that the velocity calibration does not depend on the distance of the PN from the center of the galaxy, the $x,y$ position of the PN on the CCD or the velocity of the PN. 

\begin{figure}
\includegraphics[width=0.35\textwidth, angle=270]{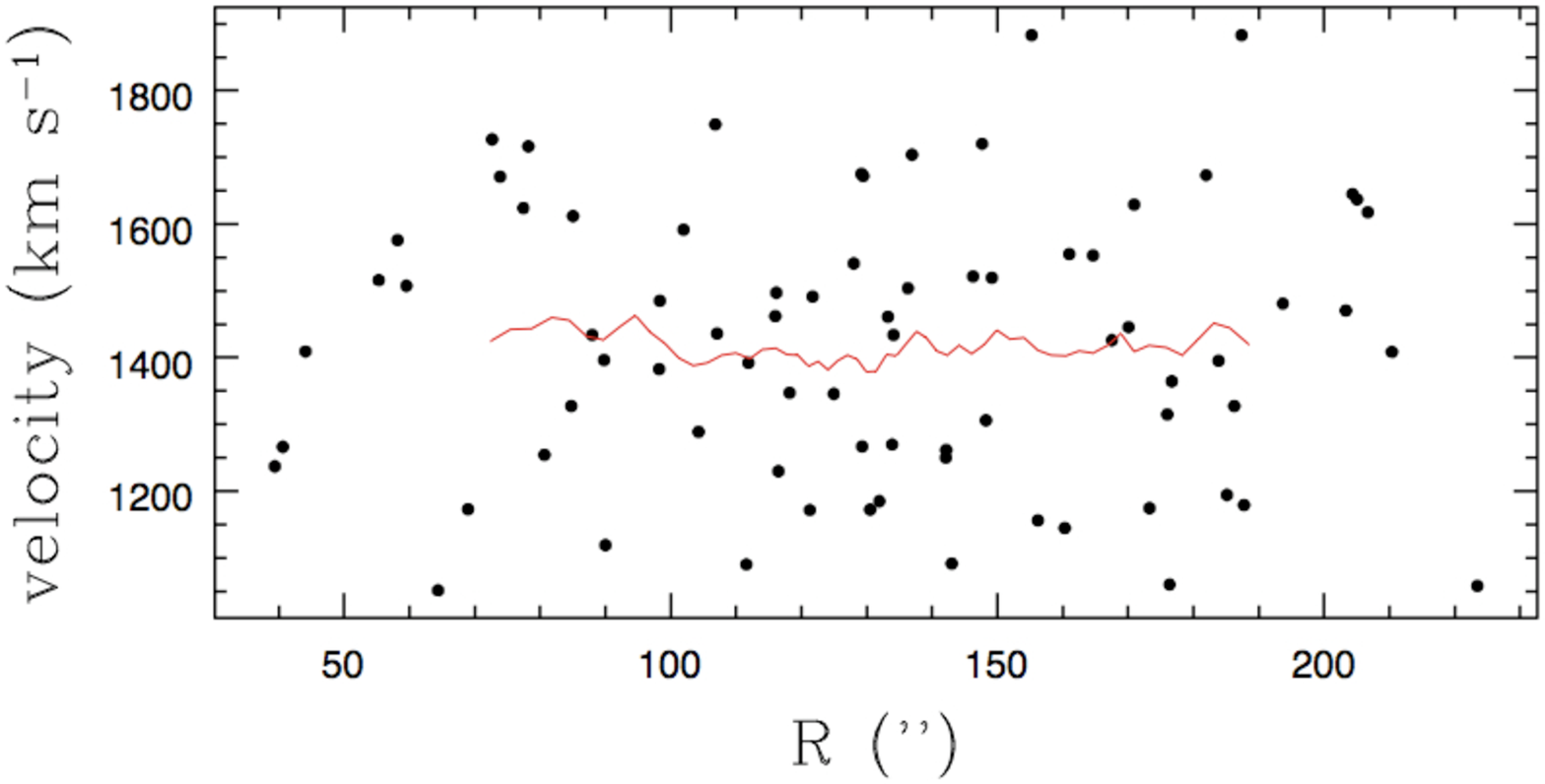}
\caption{ \label{runavpic} Velocity distribution of our central field's PN candidates as a function of radius. The red line marks the running average computed with 20 PNe in an annular bin. The lack of a trend in the average indicates that the velocity calibration does not vary with radius. }
\end{figure}

In the first test, we computed a running average of the velocity as a function of radius for the PNe in the central field (field 5). Assuming point symmetry, we expect to see a flat mean-velocity curve since the running average relies on averaged annular bins.  Since it is not physically possible for the annular bins to have a velocity gradient, a rising or falling curve would indicate that our velocity calibration depends incorrectly on radius.  In Fig.~\ref{runavpic}, we observe a flat running average, which indicates that our velocity calibration is not dependent on the distance to the center of the galaxy.

\begin{figure}
\includegraphics[width=0.45\textwidth]{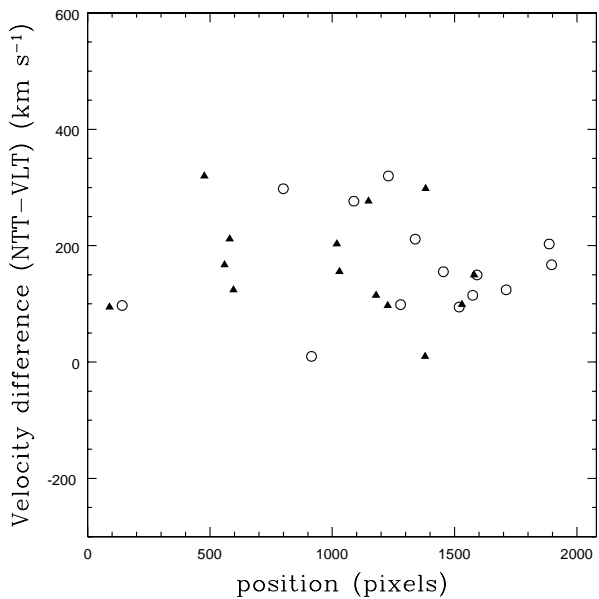}
\caption{ \label{xandy} 14 PNe in common between our sample and the 1994 NTT sample as a function of position on the CCD. The triangles represent the difference in velocity as a function of $x$-position and the open circles show it as a function of $y$-position. The lack of a trend in either distribution indicates that the velocity calibration is not a function of $x$ or $y$ on the CCD. The mean difference in velocity is 166 km s$^{-1}$ with $\sigma$=88 km s$^{-1}$. These values are addressed in \S \ref{sysoff}. }
\end{figure}

In the second test we looked for a trend in the velocity as a function of position on the CCD. We compared our velocities with those from \cite{Arnaboldi:1994a} using 14 objects in common between the two samples with a mean difference of 166 km s$^{-1}$. We show the dependence in the difference as a function of $x$ and $y$ in Fig.~\ref{xandy}, and we do not see a scaling with position. 

\begin{figure}
\includegraphics[width=0.45\textwidth]{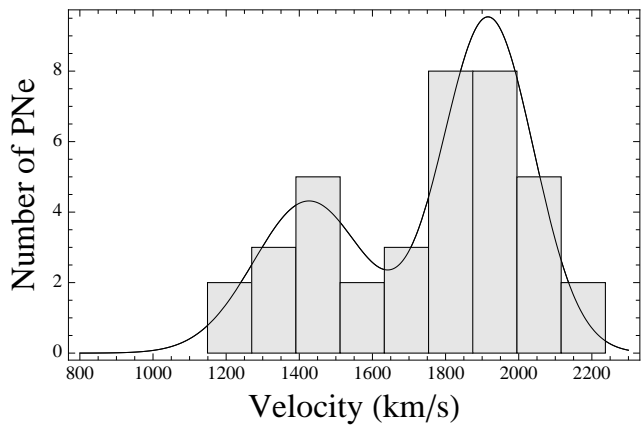}
\caption{ \label{difff7} Histogram of velocities in field 7 is overplotted with the modeled velocity distribution.  The model is a maximum-likelihood fit to the unbinned velocity distribution. We can compare the separation between the peaks to the difference in the published systemic velocities of NGC 1399 and NGC 1404. In this figure, we have applied the systematic shift of 166 km s$^{-1}$ discussed in \S \ref{sysoff}. After the shift, the fitted galaxy velocities are 1426 km s$^{-1}$ and 1917 km s$^{-1}$.}
\end{figure}

The final test is the measured difference between the mean velocity of NGC 1399 PNe and the mean of the NGC 1404 candidates that we can see in field 7. If our velocity calibration were scaled with velocity, we would expect to see the difference in the means of the two galaxies scaled proportionately. Using a maximum likelihood technique we fit the PNe in field 7 with two normal distributions, one for each galaxy. See Fig.~\ref{difff7}. 

We see a difference of 491$\pm 55$ km s$^{-1}$ between the peaks of NGC 1399  and NGC 1404. The fit shown in Fig.~\ref{difff7} finds the peaks at 1260$\pm 47$ km s$^{-1}$ and 1751$\pm$28 km s$^{-1}$. This separation is consistent with the separation of the values in the literature of 522 km s$^{-1}$ \citep{Graham:1998}. The spatial selection caused by the boundaries of field 7 may influence the mean velocity of the NGC 1404 objects detected. As discussed in \S \ref{1404kin}, NGC 1404 displays rotation; according to \citet{Graham:1998} it would have the effect of lowering the velocities that we detect in NGC 1404 which would account for the slightly smaller measured separation.  Our conclusion is that  the velocity calibration does not change as a function of distance to the center of the galaxy, position in the field or velocity. We conclude that we are accurately measuring relative velocities with  this technique. We discuss the determination of the absolute velocity in the next section. 

\subsection{Calculating absolute velocities}
\label{sysoff}
As described in \S \ref{caveat}, the measured instrumental velocities are relative velocities. Absolute velocities are necessary to compare these measurements with other kinematic tracers. The mean velocity of the current sample is 1197 km s$^{-1}$. Using a friendless outlier algorithm, we remove 12 low-velocity planetary nebulae that we believe are not virialized and therefore do not trace the mass of NGC 1399 (See \S \ref{lowvelpop}). The mean instrumental velocity for the 146 PNe associated with NGC 1399 is 1250 $\pm 17$ km s$^{-1}$.

\begin{figure}
\includegraphics[width=0.35\textwidth, angle=270]{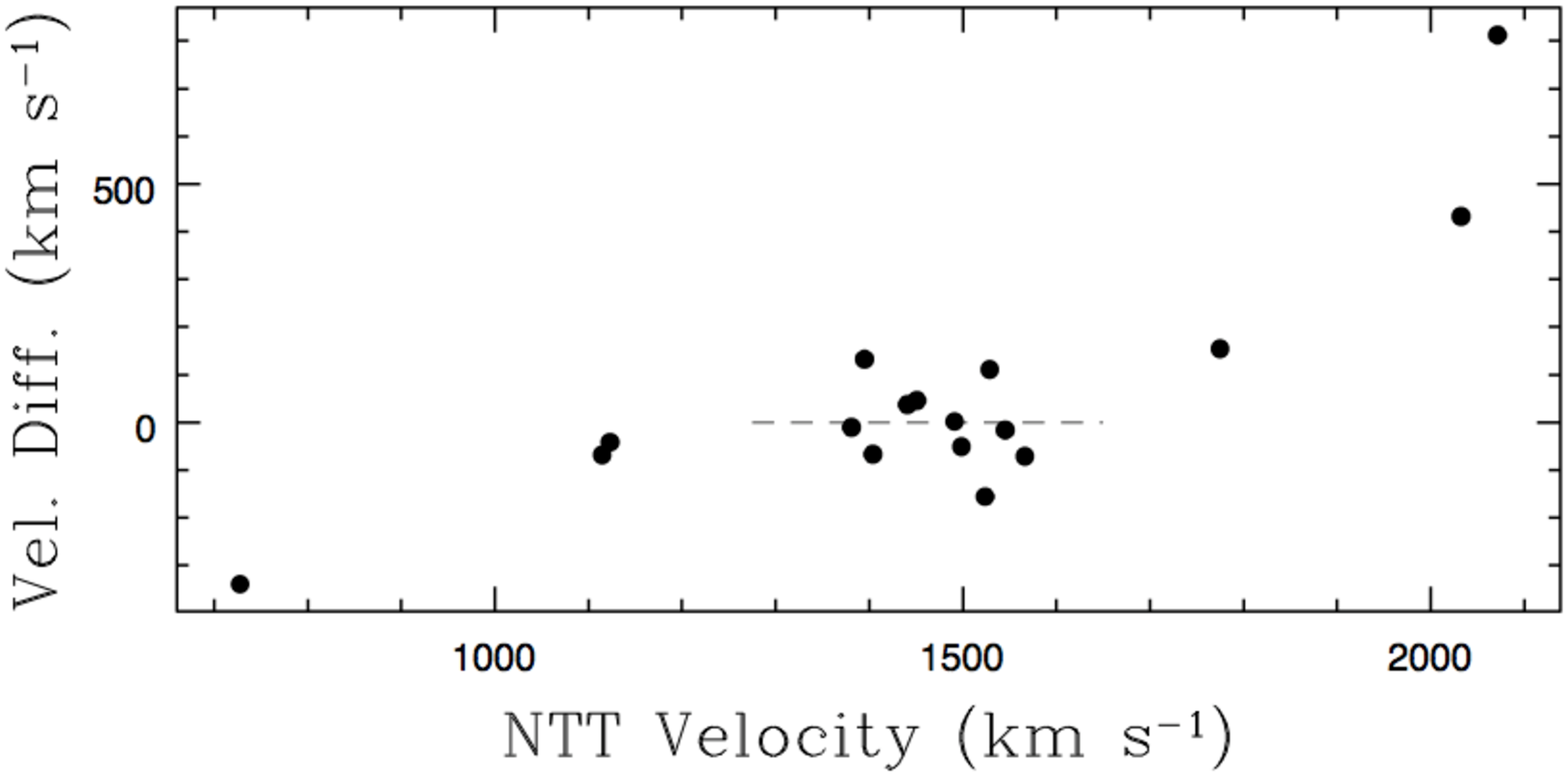}
\caption{ \label{ntt} Comparison of the current velocity measurements with those from the NTT measurements (\cite{Arnaboldi:1994a}). These 17 objects were detected in both samples. For the middle range of NTT values, we see a consistent agreement with the more recent data. At the fast and slow extremes, the agreement deviates. In this figure, we have already applied an offset of 166 km s$^{-1}$, which is the mean offset in the 14 PNe at the center of the distribution  (shown with the dashed line).}
\end{figure}

We compare the VLT counter-dispersed-technique velocities with the 1994 NTT multi-object-spectroscopy measurements \citep{Arnaboldi:1994a}. As shown in Fig.~\ref{ntt}, a comparison indicates that the offset is constant for objects in the center of the NTT velocity distribution.   Relying on the 14 out of 17 objects that were detected in both samples in this range, we calculate a systematic offset of 166 km s$^{-1}$.  The dispersion in the difference, 88 km s$^{-1}$, is consistent with the combination in quadrature of the errors in the VLT and NTT samples: i.e. $37^2 + 72^2 \simeq 88^2$. As the distribution in Fig.~\ref{xandy} illustrates, the offset between the previous work and this velocity calculation is not dependent on position on the CCD.

The error in the offset is calculated from the dispersion, $88/\sqrt{14} = 24 $km s$^{-1}$.  When we apply the shift with this error to our measured values, we find a mean velocity of 1416$\pm29 $km s$^{-1} $ for our sample of PNe in NGC 1399. This is consistent with \citet{Graham:1998} where v$_{sys}$=1425$\pm$9 km s$^{-1}$ (NED) and \citet{Richtler:2004} where v$_{sys}$= 1441km s$^{-1}$ for the GCs. Since the offset is constant, our measurements stand alone as relative velocities. For absolute velocities, we rely upon the adjustment of 166 km s$^{-1}$ based on the NTT values.

\subsubsection{Comparison to X-Shooter} \label{xshooter1} 

X-Shooter is a new, multi-wavelenth, medium-resolution, single-target spectrograph for the VLT \citep{DOdorico:2008}. It is designed to cover a wavelength range of $\lambda$3000-24800 using three arms: UVB ($\lambda$3000-5595), VIS ($\lambda$5595-10240) and NIR ($\lambda$10240-24800). As part of the commissioning observations, we received observations of one of our PNe from the UVB and VIS arms. We were able to resolve the double [OIII] emission at $\lambda$4958.9 and $\lambda$5006.8. The long wavelength range and a multitude of skylines gives us a wavelength calibration with 1\AA\ precision and yielding a measured velocity for this PN of $1117\pm60$ km s$^{-1}$.    Having accounted for the systematic offset of 166 km s$^{-1}$ discussed in \S \ref{sysoff}, this value is consistent with the measurement of $1059\pm37$ km s$^{-1}$ from the counter-dispersed images.

\subsubsection{Understanding the velocity offset}
\label{veloffset}

In \S \ref{caveat} we demonstrated that the velocity calibration is internally self-consistent. Now we discuss the systematic offset. Relative to the measurements  from \citet{Arnaboldi:1994a}, we see a difference of 166 km s$^{-1}$. We explored several possibilities for the origin of this offset.

The first possibility questioned the assumption that the short stellar spectra are symmetrical in wavelength. Depending on the temperature of the star, it seemed possible that metal lines were creating an asymmetrical feature on the blue side of the stellar spectrum as seen through the filter. Our assumption of symmetry is important because we are using the locations of stars in W0 and E180$_{rot}$ to register the images. If asymmetry caused a systematic shift of the center of the filter bandpass by say 2 pixels, it would introduce an error in the mean velocity of 131km s$^{-1}$. 

\begin{figure}
\includegraphics[width=0.45\textwidth, angle=90]{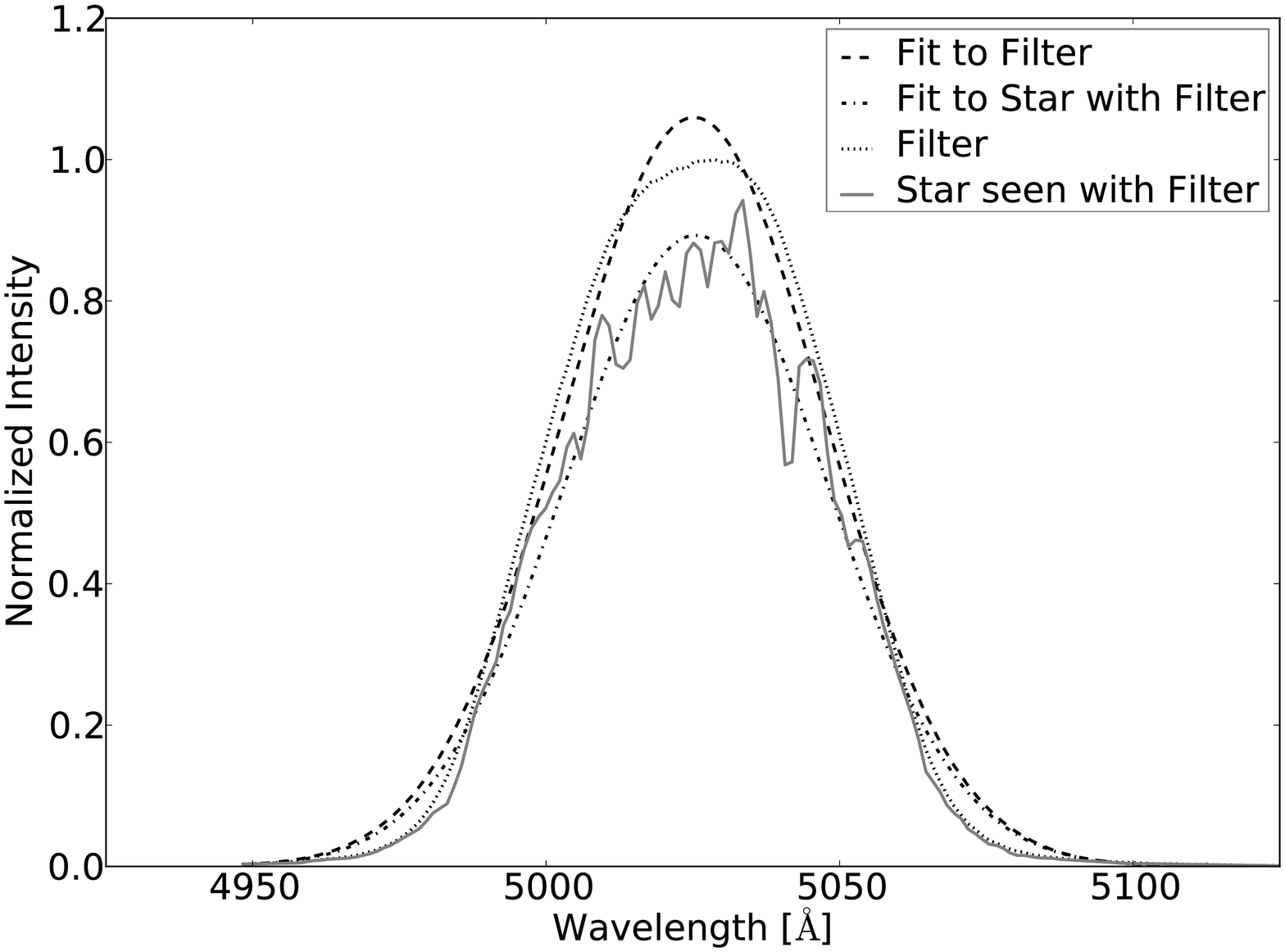}
\caption{ \label{stellar.asym} Comparison of the filter to a stellar template as seen through the filter. The filter response curve is shown as the dotted line and fit with a dashed Gaussian. The solid, gray line represents a 5500K [Fe/H]=0 stellar template \citep{Munari:2005} as seen through the filter. Notice the asymmetry caused by metal lines in the blue end of the spectrum. Comparing the dash-dotted Gaussian stellar fit to the dashed fit to the filter, we see that the measured position of the center will not be significantly affected by the absorption ($<0.3$ \AA ). }
\end{figure}

This issue was resolved by comparing the measured center of the filter to the measured center of a stellar template point-by-point multiplied with the filter bandpass. The centers were obtained by fitting Gaussians. The stellar templates are ATLAS9 model atmospheres with log($g$)=4.5, no rotation and solar metallicity \citep{Munari:2005}. We covered a range of 4000-8000K. A typical fit to the filter and the filter-template combination is shown in Fig.~\ref{stellar.asym}. We conclude that, over this range of temperatures, the fitted center of the stellar spectrum compared to the Gaussian fit to the filter differs by less than 0.3 \AA \ or 20 km s$^{-1}$ and therefore  the absorption lines are not the cause of the zero-point offset. 

The second possibility for the cause of the velocity offset is the influence of temperature on the filter bandpass. Interference filters in convergent beams generally experience a shift of 1\AA \ to the blue for every 5$^{\circ}$C drop in ambient temperature \citep{Jacoby:1989a}. Since the calibration frames are taken during the day, they could have introduced a systematic shift. However, according to the condition report, the dome was climate controlled to the night temperature during the day. We conclude that the temperature did not affect the velocity measurement.

After completing the astrometry and confirming its accuracy to 0\arcsec.3 in the dispersion direction (see \S \ref{astrometry}), we recalculated the velocities using celestial coordinates. This approach cuts out the registration of the images since the astrometry is completed independently for the two frames. Additionally, the stellar astrometry confirms that we understand the geometry of each plane (W0 and E180) independently. For each PN, we obtained $\alpha,\delta$ (J2000) in the W0 and E180 frames. After calculating the separation in arcseconds, we used the plate scale (0.25\arcsec/pix) from the astrometry to convert to a separation in pixels and subsequently \AA ngstroms using the local dispersion ($\sim$1.2 \AA/pix). The technique did not improve the agreement between our measurements and the 1994 NTT values. 

In conclusion, we have examined several sources for the velocity offset and none of them are viable. We are unable to account for this difference between our measurements and the literature. Because we are confident that our velocities are internally consistent, we approach them as accurate relative velocities.

\subsection{Mosaicking the individual pointings on a global velocity field}
\label{lock}

\begin{table}
\caption{ Duplicated detections from regions of overlap between fields }
\label{duptab}
\centering
\begin{tabular}{cccc}
\hline \hline
 Field & Field & \# of Duplicates & Velocity offset$^a$\\
\hline
 2 & 5 & 3 & 18 km s$^{-1}$  \\
 4 & 5 & 10 & -24 km s$^{-1}$ \\
 7 & 8 & 3 & -55km s$^{-1}$ \\
 4 & 7 & 1& 22 km s$^{-1}$ \\
 8 & phantom PNe & 16 &47 km s$^{-1}$\\
\hline
\end{tabular}
\begin{list}{}{}
\item[$^{\mathrm{a}}$] The velocity offset is the amount necessary to add to the velocities  from the first column in order to make them consistent with the velocities from the field in the second column
\end{list}
\end{table}

Some of the fields have regions of overlap, and there are 17 PNe detected in these regions. We use these duplicated detections to try to understand if the systematic velocity offset is global or if it changes field to field. We compared velocity measurements from each overlap region to determine the offset necessary to make the two fields in question consistent. 

With the central field (field 5) defined as our benchmark, we adjusted two fields (field 2 and field 4) using PNe duplicated in field 5. We applied a high-sigma criterion for rejecting the outliers. Field 2, the topmost field, contained only 3 usable PN duplicates for this purpose, and we calculated an offset of 18 km s$^{-1}$ between the objects in field 2 and field 5.  Field 4, just to the east of the central field, contains 10 duplicates, and we calculated an offset of -24 km s$^{-1}$ between the objects in field 4 and field 5. We apply the calculated difference to each field creating a global velocity field from the three local ones. Since these adjustments are on the order of the error in the velocity measurement, we find these three fields consistent with a global systematic offset. The duplicates are summarized in Table~\ref{duptab}.

The two southernmost fields do not contain reliable overlap with the fields above (see Fig.~\ref{grid}). We created a smoothed, symmetrized velocity field using the now-consistent velocities from the three known fields (see \S\ref{rotation} regarding the construction of this velocity fields). By assuming point symmetry of the PNe in fields 2, 4, and 5 and smoothing, we create a velocity field that covers the southern pointing (field 8). The field was evaluated at the location of 16 observed field 8 PN candidates thereby creating an equal number of phantom PNe. We then adjusted the mean velocity of the field 8 candidates to match the mean velocities of our phantom PNe by adding 47 km s$^{-1}$. Now, having made the field 8 velocities consistent with the rest of the field, we used 3 duplicate PNe to calibrate the velocities of the field 7 candidates, adding 55 km s$^{-1}$ to bring them onto the same plane as fields 2, 4, 5, and 8. There is one PN in common between field 4 and field 7, but we favored the more robust adjustment to field 8 instead. This procedure ensures that the velocity measurements in all of our fields are self-consistent.

\subsection{The separation of NGC 1399 and NGC 1404}
\label{separation}
As shown in Fig.~\ref{grid}, the central part of NGC 1404 is observed in the southeast field (field 7). The PNe in this field come from both NGC 1399 and NGC 1404.  The published mean velocities of these galaxies differ by over 500 km s$^{-1}$, but, due to the dispersions, it isn't possible to separate the PNe unambiguously by their velocities. We have adopted a probability of membership method that assigns each PN to a galaxy only if it has 95\% probability of belonging. 

The first criterion of membership is the relative surface brightness contribution from the two galaxies at the location of the PN.  The light profile of NGC 1399 is taken from Dirsch et al (2003). The light profile for NGC 1404 is an $r^{1/4}$-law fit to photometry data from the same paper. The location of the planetary was then projected onto the major axis of each galaxy and both light profiles were evaluated at the respective radii. The ratio of flux at the projected distance of the PN is the ratio of probability of membership based on light alone.

The second criterion is the relative likelihood that an object of a given velocity belongs to the velocity distribution of NGC 1399 versus NGC 1404. The total velocity distribution of field 7 was fitted with two Gaussians through a maximum likelihood technique (Fig.~\ref{difff7}). Then, reversing the technique, the relative likelihood of membership for each velocity was calculated. The ratio of likelihoods provides a ratio of probability of membership from velocity only.

Our two criteria were multiplied to produce a ratio of probability of membership value.  A planetary is assigned membership if it is 19 times more likely to belong to one galaxy than the other (95\% confidence interval). The 95\% confidence criterion results in only 6 objects of ambiguous membership. If we lower our standard to 60\% confidence of membership we would have 2 ambiguous PNe--- 2 more PNe would belong to NGC 1404 and 2 more to NGC 1399. We retain the more stringent criteria. In order to examine the specific cases of the ambiguous PNe, see the figures in \S \ref{phasespacediag}.  The Field 7 candidates contribute 9 members of NGC 1399, 23 members of NGC 1404, and 6 of ambiguous membership. This analysis results in a total of 146 NGC 1399 planetary nebulae in our sample.


\section{Astrometry}
\label{astrometry}
The astrometry was performed using the \texttt{imcoords} package in IRAF. We visually matched stars from USNO-B \citep{Monet:2003}  to those in our images (about 10 per field). Because they are spectral images, for each star we fitted a 2-dimensional Gaussian and shifted the location to account for the local bandpass shift before building up a matrix of star positions. It is not necessary to correct for the anamorphic distortion because that transformation is folded into the mapping from the spectral plane to the celestial plane. Our images are described with a 2nd-order solution relative to the USNO-B stars. We expect our absolute astrometry to be no better than the errors on the USNO-B coordinates which is $0''.2$ \citep{Monet:2003}. 

To check the validity of our astrometry we selected several stars that were not used in the astrometric fit and compared their calculated VLT (i.e. this sample) positions to the USNO-B coordinates. The residuals are $<$0\arcsec.4. We also completed an astrometric plate solution for a narrow-band NTT image covering roughly the same field. When we compare the NTT astrometry to the USNO coordinates, the residuals are 0\arcsec.3--- comparable to those for the VLT  data. This indicates that the spectral imagery of the VLT data does not cause an unpredictable behavior in the geometry that would affect our velocity calibration.

For each planetary nebula, we average the position in the W0 and E180$_{rot}$ frames to get velocity-independent position coordinates. We applied the astrometric solution to these coordinates yielding an  $\alpha,\delta$ (J2000) catalog for the emission-line objects. This catalog will be included in a subsequent paper. 

\subsection{Comparison to X-Shooter} As discussed in \S \ref{xshooter1}, a PN from our sample was included in the commissioning data for the UVB and VIS arms of X-Shooter. The pointing coordinates for the PN were taken from the astrometric solution from the counter-dispersed images. The slit placement required a blind offset because PNe would not be visible in an acquisition exposure. As expected, the PN lands 1/3 of the way along the 10\arcsec--long slit. It appears very bright, so we assume that the object is well-centered on the slit. The X-Shooter observations indicate that our absolute astrometric solution is near the limit imposed by the USNO-B coordinates.


\section{The PN velocity sample}
\subsection{Phase-space diagrams}
\label{phasespacediag}

\begin{figure}
\includegraphics[width=0.45\textwidth]{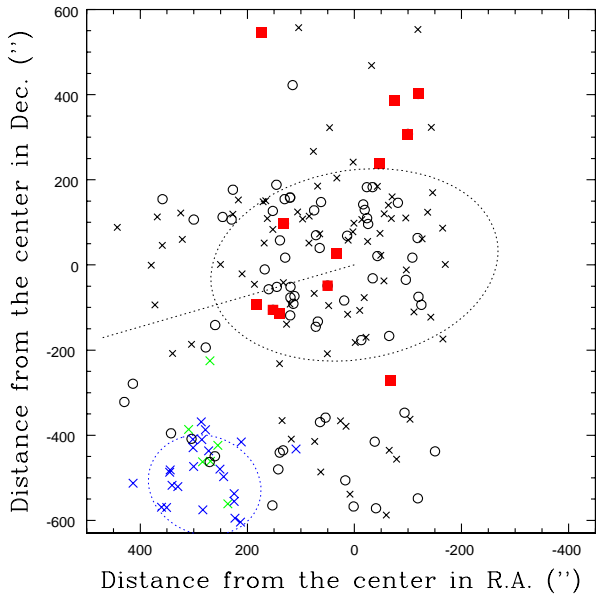}
\caption{ \label{space} The locations of the PNe in space. PNe moving faster than the systemic velocity are marked with black crosses. Those moving slower than the systemic velocity of NGC 1399 are marked with black circles. The low-velocity objects are marked with solid red squares. Blue crosses indicate objects associated with NGC 1404 and green crosses mark the ambiguous PNe that are not assigned to either galaxy. North is up and the major axis lies at 110$^{\circ}$ (dotted line).  The dotted ellipses represent the C=25mag/arcsec$^2$ isophotes based on \citet{Dirsch:2003}.}
\end{figure}

 In this section we view the PN 2-D velocity field in various projections in phase space in order to investigate the kinematic properties of the galaxy.  In particular, we present our PN sample in four subcomponents: those associated with NGC 1399, those associated with NGC 1404, those ambiguous ones that have not been assigned to a galaxy and a low-velocity subcomponent which we believe is not bound to either galaxy.  The NGC 1404 and ambiguous candidates are defined in \S \ref{separation} and the low-velocity subcomponent is defined in \S \ref{lowvelpop}. The spatial locations of the planetary nebulae  are shown in Fig.~\ref{space}. Notice that the positions are primarily limited by the locations of the five pointings shown in Fig.~\ref{grid}. 

\begin{figure}
\includegraphics[width=0.45\textwidth]{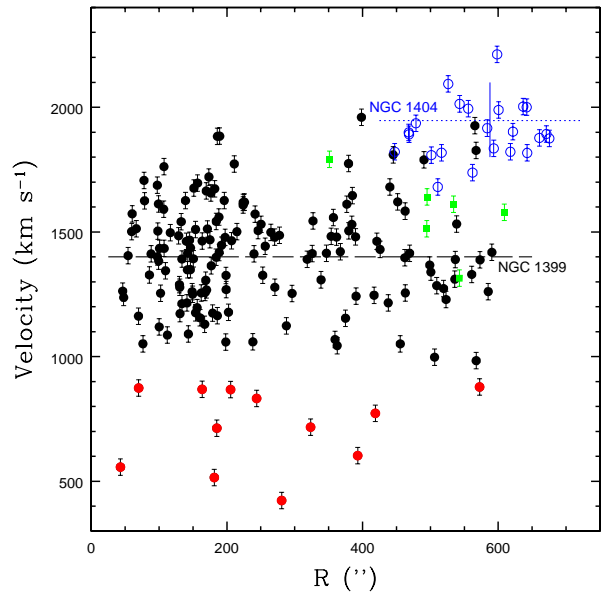}
\caption{ \label{phase} Phase space distribution. The NGC 1399 planetary nebulae velocities are shown as a function of radius in black. The 12 low-velocity objects are marked in red.  The open blue circles show the NGC 1404 PNe and the green squares are those of ambiguous membership (see \S\ref{separation}). The systemic velocities of the two galaxies are marked in dashed (NGC 1399) and dotted (NGC 1404) lines. The position of NGC 1404 is marked with a solid blue line at r=588\arcsec. Notice that a PN can be assigned to NGC 1399 based on its P.A. despite being at a radius and velocity consistent with NGC 1404.  }
\end{figure}

Comparing the line-of-sight velocity to the projected radius in Fig.~\ref{phase}, we see the main NGC 1399 population in black with the measured systemic velocity of this subcomponent marked with the black dashed line. Notice the paucity of PNe at 300\arcsec--- this is caused by the spatial coverage of the survey.  There is a low-velocity subcomponent moving at about 700 km s$^{-1}$ slower than the galaxy. Referring back to Fig.~\ref{space}, these objects are distributed across the north-central regions of our observations. As confirmed in Fig.~\ref{runavpic}, the average velocity of the main population is constant at all radii. 

\subsection{PN kinematics of NGC 1404}
\label{1404kin}
The PNe assigned to NGC 1404 (marked in blue in Fig.~\ref{space}) dominate the sample in the southwest. In Fig.~\ref{phase}, these objects are found at large radii, around the systemic velocity of NGC 1404 (1947 km s$^{-1}$; \cite{Graham:1998}). As illustrated in Fig.~\ref{difff7}, the fitted mean velocity for the NGC 1404 PNe in this sample is 1917 $\pm{ 37}$ km s$^{-1}$. \citet{Graham:1998} report rotation in NGC 1404 that could cause the PNe that we are detecting to be up to 100 km s$^{-1}$ slower than the systemic velocity of NGC 1404. This may contribute to the $\sim$30 km s$^{-1}$ difference between the fit shown in Fig.~\ref{difff7} and the mean velocity value in the literature. 

\subsection{Low-velocity subpopulation}
\label{lowvelpop}
We selected the 12 low-velocity objects from our sample using a high-sigma clipping algorithm similar to the technique in \cite{Merrett:2006}.  It is an iterative outlier test that rejects objects that lie $>2.75\sigma$ from the mean of the total sample.  The clipping limit of $2.75\sigma$ is chosen because 99.3\% of a normal distribution lies within $2.75\sigma$. In a sample of 146 objects, we expect to find less than one object outside this range.  This subpopulation of 12 objects  has a mean velocity of 718 km s$^{-1}$.   Removing these low-velocity outliers symmetrizes the sample around NGC 1399's ${\rm v}_{sys}$ because there is no corresponding high-velocity counterpart. 

In the high-sigma clipping algorithm, we assign PNe membership in either NGC 1399 or the low-velocity subcomponent. It is possible or even likely that one or two of the objects assigned to the subcomponent are actually members of NGC 1399. However, in most of the analysis in this paper, it is important to define the sample as a discrete set of objects, and the results will not change significantly based on the misclassification of one or two objects. This is not the case for the calculation of the velocity dispersion, which is sensitive to outliers in the wings of the distribution. The velocity dispersion calculation employs a robust fitting technique which is detailed in \S \ref{robfitting}.

\begin{figure}
\includegraphics[width=0.45\textwidth]{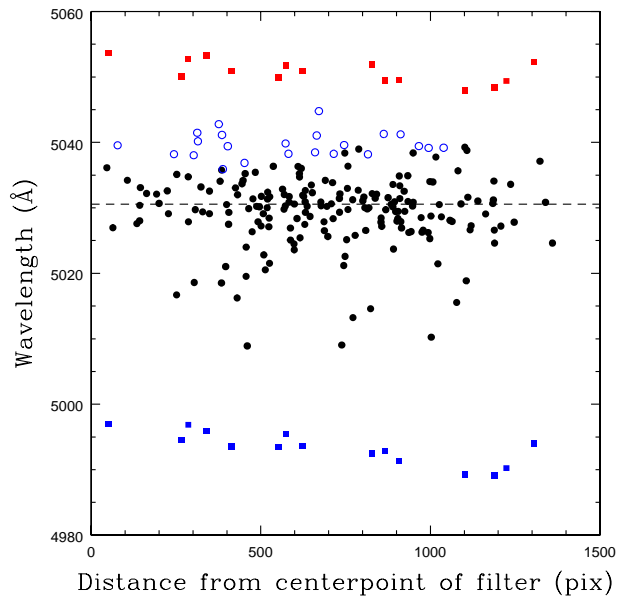}
\caption{ \label{filtsym} The detection limits imposed by the filter. This figure shows all of the PNe  detected in our five fields. The black circles represent the ones associated with NGC 1399, the low-velocity subcomponent and the unassigned PNe. NGC 1404 PNe are marked with a blue open circle. The dotted line marks the systemic velocity of NGC 1399  \citep{Graham:1998}.  The red and blue squares trace the red and blue 50\% transmission limits of the filter respectively. At small distances, the mean wavelength is near the mean wavelength of the galaxy. As the bandpass shifts at the edges of the CCD, the sample gets closer to the edge of the filter. Since there are no detections at or near the edge of the bandpass, we conclude the filter is not blocking the high-velocity counterpart to the low-velocity subpopulation. Moreover, we are detecting NGC 1404 PNe at these high velocities--- it would be possible to see NGC 1399 ones at the same wavelength if they existed.}
\end{figure}

The low-velocity objects are remarkable in that they are not uniformly distributed over the entire galaxy, but rather concentrated in the north-central parts. Considering the asymmetric placement of the fields and the relatively small numbers of PNe in some of the outer fields, we considered the possibility of a selection from the filter. Because it is an interference filter at an angle in a converging beam, the bandpass function is not centered on the CCD, as shown in Fig.~\ref{bpshift}. We examined the 50\% transmission edges of the filter as a function of distance from the intersection of the plane of the CCD with the axis of the filter, as seen in Fig.~\ref{filtsym}. We conclude that the filter is not causing the asymmetry in the velocity distribution. The most recent globular cluster (GC) work also shows a handful of low velocity objects in the sample removed with a tracer-mass-estimator method  \citep{Schuberth:2009}, which supports our claim that we are seeing a real substructure. 

A low-velocity subcomponent is difficult to detect with integrated light measurements. Based on the phase-space distribution of the low-velocity PNe, we can infer that the subcomponent would manifest itself as an asymmetric feature in the line profile. This retrograde tail would be visible in the h3 value from integrated light measurements. \cite{Saglia:2000} present h3 values of about -0.1 at large radii along the minor-axis slit. These measurements agree with the asymmetry in the PN distribution tending towards low velocities, but they do not provide an estimate for the amount of light or the mean velocity of the subcomponent.

The origin of the low-velocity subcomponent is unknown. The mean velocity is similar to the systemic velocity of NGC 1396 (808$\pm 22$ km s$^{-1}$, \cite{Drinkwater:2001}), an elliptical galaxy to the west of NGC 1399, and their spatial distribution is consistent with their association with the outer halo of this galaxy. Unfortunately, our observations do not extend to the west, so we lack more conclusive evidence.  The low-velocity globular clusters are dominated by blue metal-poor members. This is consistent with the possibility of the low-velocity material originating from a stripped galaxy. We do not detect a velocity gradient in the low-velocity PNe, so it seems unlikely that these objects are being currently stripped as part of a interaction between the two galaxies. The photometry of NGC 1399 from the OBEY survey \citep{Tal:2009} indicates a smooth light distribution down to the surface brightness limit of V=27.7 mag arcsec${^{-2}}$, so the stellar component accompanying the low-velocity PNe must be extremely diffuse across our fields. This corroborates the indication from the velocity gradient that we are not seeing a clearly defined stream of PNe from an infalling galaxy.

In any case, the asymmetry in the velocity distribution indicates that the low-velocity PNe are not currently phase-mixed with NGC 1399. Like in M87 \citep{Doherty:2009}, we see a superposition of kinematic components in the halo of this central cluster galaxy. This reflects the complexity of galaxy superpositions and interactions in cluster cores.  The subcomponents could not have been disentangled without discrete velocity tracers; having PN velocities is therefore important for deriving a complete picture of  the kinematics in the outer halos of brightest cluster galaxies.  The detection of subcomponents in galaxy halos is not restricted to cD galaxies. In cases like M31 or the MW, these structures can be traced with red giant branch (RGB) stars. In nearby clusters, we are unable to detect individual RGB stars, but it is possible to measure the velocities of individual PNe. Globular cluster velocities can also be used outside the local group, but we emphasize that PNe hold an important role because they are the only individual stars visible at several Mpc and they trace the stellar population directly.

\section{Kinematics of the halo of NGC 1399}
\subsection{Rotation}
\label{rotation}
\begin{figure}
\includegraphics[width=0.45\textwidth]{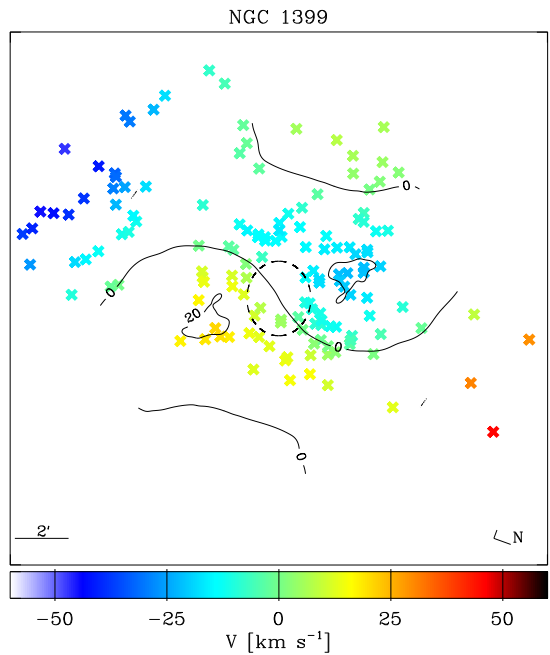}
\caption{ \label{smoothmap} NGC 1399 halo velocities. This smoothed, symmetrized velocity map shows the location of each PN with an x. The photometric major axis is vertical in this configuration. The map indicates rotation of $\ll{50}$ km s$^{-1}$. The kinematic major axis is offset from the photometric major axis-- the kinematic axis twists at large radii. In this figure, the 12 low-velocity PN were not included since we assume that they are not part of the dynamical system.  The dotted ellipse represents 2R$_e$ \citep{Saglia:2000}. }
\end{figure}

Now we consider the kinematic properties of the main NGC 1399 sample. The smoothed, symmetrized velocity map shown in Fig.~\ref{smoothmap} illustrates the rotation in the observed fields. It is made by reflecting the entire sample (defined as the PNe which are not classified as members of NGC 1404, the low-velocity subcomponent or the ambiguous objects)  in $x, y$ and velocity and then smoothing using an adaptive Gaussian kernel smoothing  as described in \citet{Coccato:2009}. The velocity field was fit with a cosine function in order to derive the direction and amplitude of rotation as a function of radius. The measured values for these quantities and their errors (computed by means of Monte Carlo simulations) are obtained using the same process described in  \citet{Coccato:2009} and are listed in Table~\ref{rotbreakdown}. We observe rotation of $\sim 20$km s$^{-1}$ along an axis offset from the photometric major axis (P.A.=110$^{\circ}$). Additionally, the kinematic major axis displays a strong twist in the outer regions similar to those described in \citet{Coccato:2009}. 

This analysis is consistent with the low-amplitude of rotation ($<30$ km s$^{-1}$ along the major axis) observed by Saglia et al. (2000) inside 1\arcmin.  In contrast to the rotation result from \cite{Arnaboldi:1994a}, we see only a low amplitude of rotation in the outer parts---  rotation of $\leq{20}$ km s$^{-1}$  is a fraction of their value. A comparison to the globular cluster system's rotation can be found in \S \ref{gcs}.

\begin{table}
\caption{Rotation of the Halo of NGC 1399. The photometric major axis is at PA=110$^{\circ}$.}
\label{rotbreakdown}
\centering
\begin{tabular}{cccccc}
\hline \hline
Radius & \# of & Amplitude & Error in & P.A & Error\\
 &  Objects & & Amp. && in P.A. \\
\hline
$\prime\prime$ & & km s$^{-1}$ & km s$^{-1}$ & degrees & degrees \\
\hline
    60    &  22 &   15  &    7   &  233  &    28      \\
    200    &  64 &   20    &  6   &  246   &   17 \\
     400     &   61 &  20   &   5   &  339 &     20      \\ 

\hline
\end{tabular}
\end{table}

\subsection{The velocity dispersion}
\label{veldispsec}

\begin{figure*}
\includegraphics[width=0.95\textwidth]{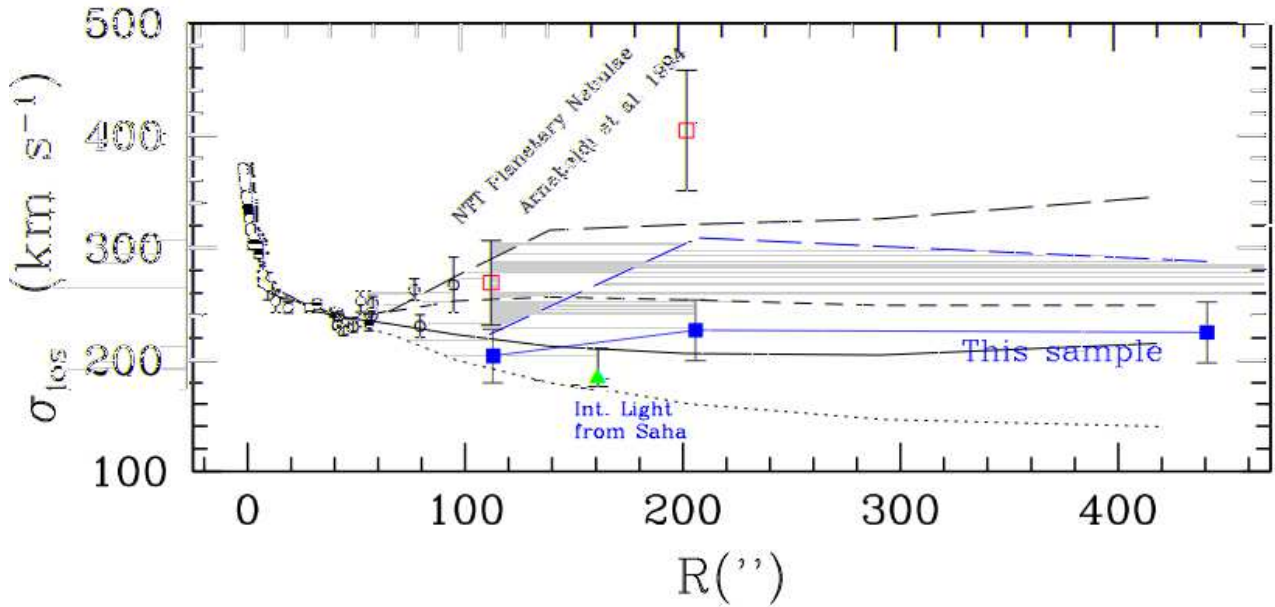}
\caption{ \label{sigma3}  Dispersion profile of NGC 1399. The new data are shown as red squares in the outer regions. The instrumental dispersion has been subtracted in quadrature. For comparison, we have shown the calculation of the dispersions without using the robust fitting (red dashed line).  The black circles represent integrated light measurements from \cite{Saglia:2000} (solid=major axis, hollow=minor axis). The minor axis is projected for comparison to the major axis. The PN measurements from \cite{Arnaboldi:1994a} are in green, rising steeply. The outermost integrated light point, a blue triangle, is from P. Saha (private communication), and it shows good agreement with a profile extrapolated from Saglia et al's (2000) major axis dispersion measurement and the new PN kinematics. The black lines are various models from \citet{Kronawitter:2000} (see discussion in text). }
\end{figure*}

Using the new PN velocities, we trace the dispersion out to  large radii. PNe are particularly good for this task because they extend well beyond the integrated-stellar-light measurements, and their number density follows the light where they overlap \citep{Coccato:2009}.  

\subsubsection{Robust Fitting}
\label{robfitting}
Due to possible contamination from the low-velocity subcomponent, the dispersion is calculated using a robust fitting technique. We begin by measuring the standard deviation of the data. To avoid using the tails of the distribution, we estimate the dispersion using only objects within 2 standard deviations ($2\sigma$) and then scale the measured dispersion using a factor generated in a Monte Carlo simulation of the same process. A 2$\sigma$ limit is selected because we are confident that the contamination is not significant within this range. A step-by-step description of the process follows in the next two paragraphs. 

Based on the number of PNe in each annular bin, 53 objects are drawn at random from a Gaussian distribution with the measured mean and dispersion of the data.  Objects at greater than 2$\sigma$ are rejected and the dispersion of the remaining objects is measured. This process is repeated 10,000 times creating a Monte Carlo distribution of dispersion values. The mean of the 10,000 dispersion values is used to calculate a scaling, $f_{2}$, between the dispersion as measured within 2$\sigma$ ($\sigma_{2}$) and the known input dispersion ($\sigma_{true}$). We define

\begin{equation}
	f_{2}=\frac{\sigma_{2}}{\sigma_{true}} .
\label{robustfit}
\end{equation}The dispersion of the 10,000 measurements gives the error in our dispersion values, $\sigma(\sigma)$. 

Returning to the data, a 2$\sigma$ limit is applied to each radial bin and the dispersion is calculated for the PNe within this limit. This dispersion is then scaled by the factor defined in Eqn. \ref{robustfit}. Since we expect the initial estimate for 2$\sigma$ to be influenced by the outliers, we repeat this process until the dispersion estimate stabilizes.  The error, $\sigma(\sigma)$, is  $f_{2} \times$ the dispersion of the Monte Carlo distribution.

Fig.~\ref{sigma3} shows the line-of-sight velocity dispersion as a function of projected radius calculated with the robust-fitting technique. The calculated values are included in Table \ref{disptable}. The radial limits of the bins are set such that all the bins have about the same number of objects (53 or 54). We discuss the dispersion, $\sigma$, rather than ${\rm v}_{rms}$, but they are approximately equivalent since the rotation velocity is low. The velocity error has been subtracted in quadrature. 

\subsubsection{Comparison to previous PN measurements}
In contrast to the previous measurements \citep{Arnaboldi:1994a}, the velocity dispersion from the new observations is much lower and much flatter. There could be several contributing factors to this change. Compared to the 1994 values, this sample is four times larger,  has more precise measurements and does not have a problem with the extreme velocities as described in Fig.~\ref{ntt}. Moreover, the robust fitting treatment has a significant effect on the dispersion. This is shown as the difference between the red squares and the red line in Fig.~\ref{sigma3}. 

In summary,  this reassessment of the velocity dispersion implies a lower value than the previously published measurements.  In the latest NGC 1399 data,  there is no evidence of an upturn in the dispersion which would have indicated a growing influence from the cluster potential.

\subsection{Other tracers}

\begin{table}
\caption{New velocity dispersion measurements in the outer halo of NGC 1399. The radius of each bin is the mean of the radii for all the objects in that bin. }
\label{disptable}
\centering
\begin{tabular}{ c c c c}
\hline \hline
 Bin & R (\arcsec) & $\sigma$ (km s$^{-1}$) & $\sigma(\sigma)$ (km s$^{-1}$) \\
\hline
\multicolumn{4}{l}{Planetary Nebulae after robust fitting}\\
\hline
 1 & 115 & 204 & 24 \\
 2 & 208 & 227 & 27 \\
 3 & 443 & 225	& 27 \\
\hline
\multicolumn{4}{l}{Revised GCs based on data from \citep{Schuberth:2009}} \\
\hline
1 &150 & 224 & 32 \\
2 & 242 & 215 & 31 \\
3 & 364 & 230 & 34 \\
4 & 508 & 214 & 31 \\
5 & 792 & 176 & 26 \\
\hline
\multicolumn{4}{l}{Integrated Light  from P. Saha (private comm.)} \\
\hline
  -- & 163 & 184 & +27/-8 \\
 \hline 
\end{tabular}
\end{table}

 \begin{figure*}
\includegraphics[width=0.95\textwidth]{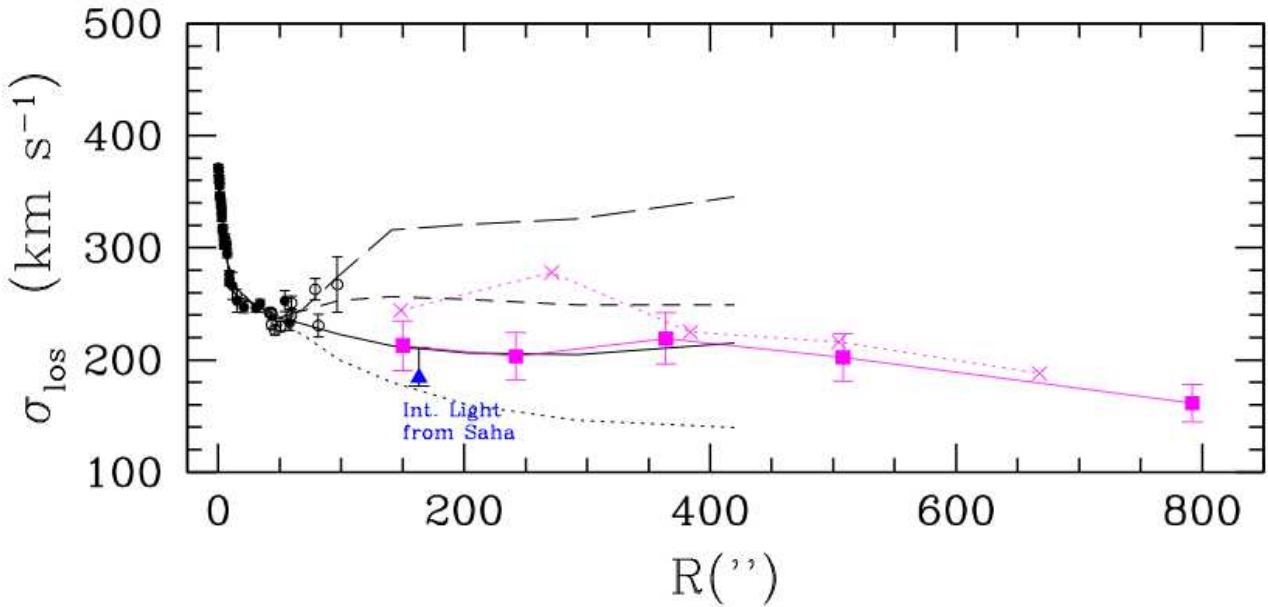}
\caption{ \label{extrasig} Globular cluster dispersion profile of NGC 1399. The black dots represent the same integrated light measurements described in Fig. \ref{sigma3}. The magenta crosses connected with a dotted line mark the GC dispersion values from \citet{Schuberth:2009}. After applying the decontamination algorithm from \S\ref{separation} to Schuberth et al.'s RI sample and also removing two low-velocity GCs based on the tracer-mass estimator method in that paper, we measure the dispersions shown in magenta squares. An instrumental dispersion of 70km s$^{-1}$ has been subtracted in quadrature. This dispersion profile is in excellent agreement with one of the models from \citet{Kronawitter:2000},  as discussed in \S\ref{modelingsec}. Lingering contamination from NGC 1404 seems to have strongly influenced the velocity dispersion profile from \citet{Schuberth:2009} around 250\arcsec. }
\end{figure*}
\subsubsection{Integrated Light}
The current PN dispersion values agree well with the major-axis values described in \citet{Saglia:2000}. One side of the parallel-to-minor-axis values does not agree well with the PN measurements, which suggests that there may be preferential contamination from non-phase-mixed PNe in this measurement. See Fig. \ref{sigma3}.

The new PN measurements agree well with the outermost dispersion measurement from integrated light--- less than 200 km s$^{-1}$ at 163\arcsec (P. Saha, private communication). This value, marked with a blue triangle in Fig. \ref{sigma3},  is derived from one all-night, carefully-sky-subtracted, long-slit observation taken with the RGO spectrograph on the Anglo-Australian Telescope. This measurement is included in Table \ref{disptable}. Approximately equal time was spent on the target and the sky.   The current PN sample is in good agreement with the largest-radius integrated light measurement confirming the consistency of these tracers. 

\subsubsection{Globular Clusters}
\label{gcs}
NGC 1399 has a rich globular cluster system with about 7000 objects \citep{Dirsch:2003}. Their abundance makes them potentially valuable tracers, but only the red subsample follows the radial distribution of the light and the PNe (see Fig. 15 in \citet{Schuberth:2009}).  We consider the velocity dispersion from the red clusters only--- the blue clusters have a higher dispersion which goes hand-in-hand with their flatter spatial distribution. Although the entire GC system is moving in the same potential as the PNe and stars, the different spatial distribution of the clusters requires different kinematics if they are to be in equilibrium.  

The GC dispersions included in \citet{Schuberth:2009} are higher than the PNe and Saha's integrated light value. These values are shown with magenta crosses in Fig. \ref{extrasig}.The difference seems to be highly influenced by contamination from NGC 1404. The decontamination in that paper is based on a cut in radius--- any objects within 3$\arcmin$ of NGC 1404 were not included. 

 We apply the two-fold decontamination algorithm (as described in \S \ref{separation}) to the red GC sample (RI in \citet{Schuberth:2009}) which leads to a sample of NGC 1399 GCs that has 10 fewer objects than RIII in \citet{Schuberth:2009}, although some of the objects that were rejected by \citet{Schuberth:2009} were kept in our sample.  We also reject two low-velocity GCs from the RI sample based on the tracer-mass estimator (TME) algorithm from \citet{Schuberth:2009}. The GCs are rebinned to have an equal number of objects in each bin, but the dispersion values fall within one error bar if we use the bin limits described in Table A.1 of \citet{Schuberth:2009}. After the decontamination, the TME outlier rejection and the subsequent rebinning, the measured dispersions drop to $\sim200$km s$^{-1}$ and are in excellent agreement with the PNe. See Fig.~\ref{sigma2}. The revised GC velocity dispersion values based on the two-fold decontamination are included in Table \ref{disptable}. 

The globular cluster rotation analysis is summarized in \S 7 of \citet{Schuberth:2009}. In agreement with the PNe, they find marginal rotation for the red sample. The red globular cluster rotation is 61$\pm35$ km s$^{-1}$ at an azimuthal angle of 154$\pm$33  degrees, but a rotation of this amplitude could be measured from randomized data with probabilty 27\% --- this is interpreted as insignificant rotation. They find rotation of 110$\pm$53 km s$^{-1}$ at $\Theta_{0}$ =130$^{\circ}\pm$24$^{\circ}$  in the blue sample, but we do not expect the blue GCs to agree with the rotation of the PNe and the integrated light based on the spatial distribution and kinematics shown in Figs. 14 \& 15 of  \citet{Schuberth:2009}.  While there is disagreement about the position angle, the planetary nebula and globular cluster kinematics agree that the amplitude of any rotation in NGC 1399 would be small.
 
\subsubsection{Fornax cluster galaxies}
Just as PNe and globular clusters can be used to trace the mass
distribution of a galaxy, cluster members allow us to estimate the
cluster's mass.  NGC 1399's position at the core of the Fornax cluster
leaves open two possibilities for the dispersion profile at extremely
large radii; the halo of a cluster's central galaxy could be
dynamically separate from the cluster, or it could be part of it. For
example, in the case of the Virgo cluster core, M87 displays a
truncation at 160 kpc, where the velocity dispersion of the galaxy
drops below 80 km s$^{-1}$ \citep{Doherty:2009}  while the
  potential determined from the X-rays is already dominated by the
  cluster mass distribution, implying that the central galaxy is
dynamically separate from the cluster.  If a central galaxy is not
truncated, then it may be difficult to define the boundary between the
galaxy and the cluster since the velocity dispersion of the galaxy
could rise with radius to become similar to that of the cluster.

\begin{figure}
\includegraphics[width=0.45\textwidth]{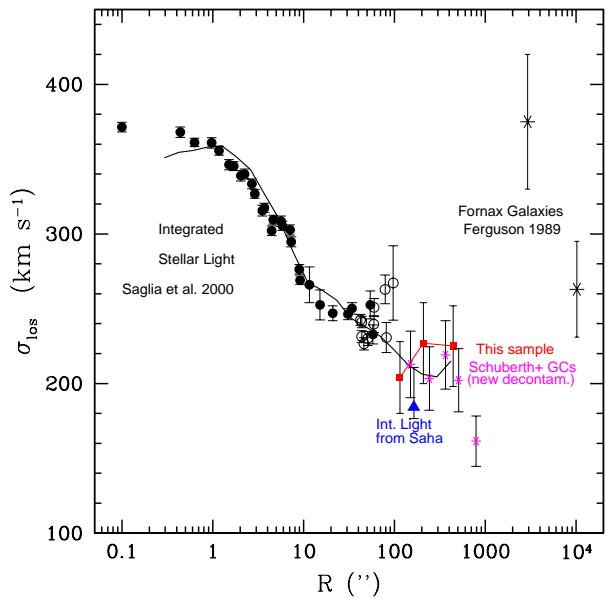}
\caption{ \label{sigma2}  Dispersion profile representing the best values for each type of tracer. As in Fig.~\ref{sigma3}, the new PNe data are shown as red squares and the black circles represent integrated-light measurements from \cite{Saglia:2000} (solid=major axis, hollow=minor axis). The last integrated-light measurement (blue triangle) agrees well with the new PNe dispersions. For comparison, the red GC sample from \citet{Schuberth:2009} has been decontaminated with the same technique as the PNe and is displayed with magenta asterisks. The black asterisks represent Fornax Cluster dispersions \citep{Ferguson:1989}. The solid line shows the best model for the PN data from \citet{Kronawitter:2000}. }
\end{figure}

 Photometry of NGC 1399 from \citet{Schombert:1986} indicates that
 there is no truncation, at least in the stellar mass, up to a radius
 of about 300kpc. The component of the Fornax cluster centered on NGC
 1399 has a dispersion of 370 km s$^{-1}$ \citep{Drinkwater:2001}. As
 discussed by Drinkwater et al., the dwarf galaxies in Fornax
 constitute a distinct population with a much higher dispersion than
 the giants (429$\pm$ 41 km s$^{-1}$ compared to 308$\pm 30$ km
 s$^{-1}$). Fig.~\ref{sigma2} shows the velocity dispersion for likely
 Fornax Cluster members taken from \cite{Ferguson:1989}. These values
 agree very well with Drinkwater et al.  The most distant dispersion
 measurements of NGC 1399 available come from the GCs-- the profile
 seems to be falling at 1000\arcsec \citep{Schuberth:2009}.  At $\sim$2000\arcsec, the dispersion profile is not well understood. \citet{Bergond:2007} see an indication of the onset of the cluster potential \citep{Bergond:2007}, but \citet{Schuberth:2008} interpret the same data as being part of the NGC 1399 system with a falling dispersion profile. Current observations do not
 bridge the space between the last GCs ($\sim$2000\arcsec) and the first
 dispersion estimate from the cluster members at nearly
 3000\arcsec. At the radial limit of our sample (500\arcsec), the PN
 dispersion reaches 220 km s$^{-1}$ indicating that these PNe are
   not in the regime dominated by the cluster potential, but still
   within the NGC 1399 halo. Mass models based on ASCA and Rosat X-ray
   data \citep{Ikebe:1996, Paolillo:2002} indicate that the transition
   to the cluster potential might be at around 500\arcsec. However,
   these mass models may not be consistent with the globular cluster
   kinematics \citep{Schuberth:2009}; this issue requires more work.

\subsection{Dynamical Models}
\label{modelingsec}
\begin{figure}
\includegraphics[width=0.45\textwidth]{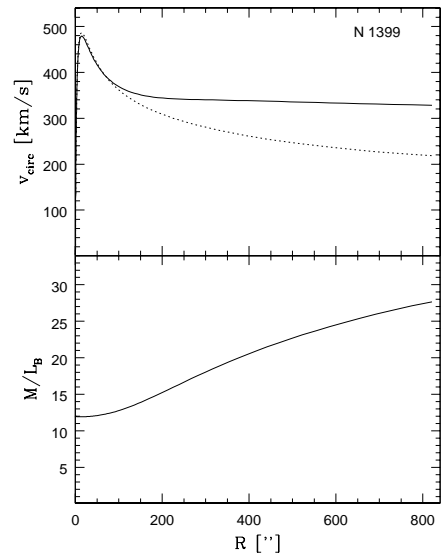}
\caption{ \label{vcmonl}  {\bf top:} Circular velocity curve for the model from \citet{Kronawitter:2000} that best fits the inner absorption line kinematics, the PN velocity dispersions and the red GC velocity dispersions (solid line). This corresponds to the model shown with a solid line in Figs. \ref{sigma3} and \ref{extrasig}. The dotted line shows the circular velocity curve of a constant mass-to-light ratio model. {\bf bottom:} B-band mass-to-light ratio profile for the best-fit model.}
\end{figure}

A full dynamical model based on the new PN and GC data is deferred to
a subsequent paper. Here we only compare the data briefly to a range
of models from \citet{Kronawitter:2000}.  These models were
  constructed based on absorption line kinematics reaching 100\arcsec,
  assuming spherical symmetry. The anisotropy of the orbit
  distribution was constrained by line profile shape measurements
  inside 80\arcsec. Best-fitting distribution functions were
  constructed in a sequence of potentials including that of the stars
  with constant mass-to-light ratio and an additional logarithmic halo
  potential. Best fit composite potentials and confidence regions were
  determined from the combined fit to all the absorption line data.
The lines overlayed on Fig.~\ref{sigma3} represent several of the
models from \citet{Kronawitter:2000};  note that that their
  projected velocity-dispersion profiles are available only to
  $R=420"$. The long-dashed model at the top represents the best fit
to the photometry and integrated-light kinematics described in
\citet{Saglia:2000}. At the bottom, the dotted line shows a
self-consistent model with constant mass-to-light ratio.  The
short-dashed line near the middle shows an intermediate model that is
consistent with the inner integrated light data as well as the
outer-most PN point. The best match to all of the velocity dispersion
points from the current PN sample and also to the globular cluster
points for $R< 400''$ is provided by  a model at the low-mass end
  of the \citet{Kronawitter:2000} confidence range, shown with a
solid line in Figs.~\ref{sigma3} \& \ref{extrasig}.   This line
  represents the velocity-dispersion profile of a mildly radially
  anisotropic ($\beta\simeq 0.25$) spherical model with a central
  B-band mass-to-light ratio of $M/L_B=12$, a dark matter halo with
  circular velocity $v_c\simeq 340$ km s$^{-1}$ for $R>200"$ and a
  mildly increasing $M/L_B$-profile, reaching $M/L_B\simeq 27$ at
  $R\simeq 800"$ (see Fig.~\ref{vcmonl}). Outside 100\arcsec \ the
  anisotropy of this model is not necessarily a good approximation for
  the halo of NGC 1399, however, because of the shallow luminosity
  density profile the velocity dispersions are insensitive to
  anisotropy \citep{Gerhard:1993, Churazov:2010}. Nonetheless, this
  model is only indicative;  more detailed dynamical modeling is
needed to explore the full range of anisotropy and mass profiles
consistent with the new data.

\section{Summary and Conclusion} 

Using counter-dispersed, slitless spectroscopy from FORS1 on the VLT,
we detected 411 emission-line objects around NGC 1399, including 146
PNe associated with NGC 1399, 23 associated with NGC 1404, 12
low-velocity PNe, 6 unassigned PNe and 224 background galaxies, and we
measured the line-of-sight velocities of the PNe. This technique,
applied for the first time on an 8-m class telescope, measures
individual velocities with errors of less than 40 km s$^{-1}$ at
nearly 20 Mpc distance. It provides an efficient way of sampling
  the velocity field in the outer halos of galaxies, particularly
because it does not require a two-stage detection-and-measurement
process. In this paper we described the reduction technique specifically developed for the astrometry and velocity calibration of these observations.  The new PN velocity measurements allow us to examine the
  kinematics of the outer halo of the cD galaxy NGC 1399.

The 146 PNe in NGC 1399 show weak rotation. The kinematic
major axis is offset from the photometric major axis and displays a
twist at large radii.  The small amplitude of rotation confirmed the
result from \citet{Saglia:2000}. The PN velocity-dispersion profile flattens at about 220 km s$^{-1}$  extending to $\sim$ 40kpc. The PNe dispersions show good agreement with the major-axis integrated-light values and the globular cluster sample \citep{Schuberth:2009} after the NGC 1404 objects have been removed using a different method. 

A comparison to the sequence of dynamical models constructed by
  \citet{Kronawitter:2000} from inner absorption line kinematics
  indicates that the PNe in the outer regions of NGC 1399 are
  consistent with a model with a B-band central mass-to-light ratio
of 12 and a dark matter halo with a circular velocity of $v_c\simeq
340$ km s$^{-1}$. This comparison highlights the importance of having
kinematics at large radii to distinguish between the various
models. This paper establishes the most up-to-date data set with
integrated light, PNe and GCs, which will constrain new dynamical
  models to be constructed in a subsequent paper.

The PN kinematics of the halo support the role of cD galaxies in
hierarchical structure formation within clusters. The multi-component
sample is comprised of objects associated with NGC 1399, NGC 1404 and
a low-velocity sub-group at $\sim$800 km s$^{-1}$.  At the
  cluster core, we expect to find a heterogenous population including
  stars left over from previous accretion, merger or tidal stripping
  events; these stars are still in the process of phase-mixing. The
superposition of the various components on the NGC 1399 field
underscores the complexity of velocity measurements in cluster cores
and the necessity of using discrete tracers to detect these
components.

 These results have opened valuable opportunities for further 
research on the formation and evolution of elliptical galaxies 
and on the dynamical structure of NGC 1399 and the Fornax cluster
core in particular. The ambiguity of the
source of the low-velocity data underlines the importance of
understanding the kinematics of the western region of NGC 1399 and the
implications on the dynamics of the Fornax cluster core. Given that
the outer regions of early-type galaxies are dark-matter dominated and
the orbits in this range have long dynamical time scales, the PN
kinematics will provide valuable constraints on dynamical studies of
this galaxy. The new data indicates that at large radii we must be
  aware of contamination in the wings of the velocity distribution as
  we look towards modeling the galaxy. Using these new PN data and
the existing GC and integrated light observations, the dynamical
modeling of NGC 1399 will be described in a future paper. Modeling NGC
1399 is a particularly exciting prospect because of existing X-ray
data providing an independent assessment of the potential. The
estimate of the orbital anisotropy will shed light on the formation of
the halos of brightest cluster galaxies. Finally, given the success of
this technique, we will apply the developed reduction processes to
other systems to further explore the kinematics of the halos of
early-type galaxies using PNe as discrete tracers.

\section*{Acknowledgments}
We thank the anonymous referees for his/her useful comments. We wish to thank R. Saglia and P. Saha for the use of their published and unpublished integrated light dispersions respectively. We thank S. D'Odorico and the X-Shooter team for the commissioning observations of one of the PNe in this sample. We thank M. Fabricius for his help with the analysis of the integrated light spectra. We are grateful to Y. Schuberth for generously sharing the globular cluster velocity catalogs.  This research has made use of the NASA/IPAC Extragalactic Database (NED) which is operated by the Jet Propulsion Laboratory, California Institute of Technology, under contract with the National Aeronautics and Space Administration.

E. McNeil acknowledges support from the Max-Planck Institute for Ex. Physics and ESO Director Discretionary Funds 2009 during a 10-month visit to the Garching campus. P. Das was supported by the DFG Cluster of Excellence. 

\bibliography{References}
\bibliographystyle{aa}

\end{document}